**E-Learning and its Socioeconomics**

*Avni Singh*

# Contents

## Acknowledgements

I wrote this paper because e-learning has always fascinated me as a topic. I have been using it extensively during the past 18 months, due to the COVID-19 pandemic. Also, over the past year, I have become involved with helping underprivileged children gain access to e-learning through my social initiative, the Society for Inclusive Education. The effects of the sudden shift to e-learning during the pandemic is something that I wanted to explore.

I want to thank Ms Ranu Sobti, the principal of Happy School, who helped me gain an insight into the lives of her students so I could provide some of them with phones for e-learning last year. Her help was paramount in establishing the Society for Inclusive Education.

Secondly, I would like to thank my parents for ensuring this paper's completion (and helping me start it). Of course, I want to also thank them for buying me books by the bulk ever since I was a child. Without that, I would not have been able to write this.

Lastly, I would like to thank Professor HS Jamadagni for guiding me through the research paper process—helping me with writing it, researching for it, and refining it. Without his help, this paper would not have been as organised or well-written.

# Abstract

While controversial, e-learning has become an essential tool for all kinds of education—especially the kindergarten-to-twelfth sector. However, pockets of this sector lack access, mainly economically underserved students. This paper explores the options available to underserved and aptly resourced members of the kindergarten-to-twelfth educational sector—a 250-million-person market, with only 9 million students enrolled in online education. The paper also provides a brief overview of the options and challenges of making e-learning available to everyone in the kindergarten-to-twelfth educational sector.

To establish whether e-learning is beneficial, it also discusses the results of a survey conducted on students and educators who have experienced e-learning, with the results showing that it is beneficial, with a general trend of teachers showing more comfort with online learning than students. The paper utilizes primary and secondary resources for this purpose, with information both from the internet, and from surveys conducted within people from the system—parents, students, and teachers.

# 1. Introduction

E-learning has revolutionised the education sector, especially with the rise of the COVID-19 virus. To uphold social distancing rules, many schools replaced offline school with platforms like Zoom or Microsoft Teams. Not only this, but supplementary education also moved online, with both paid and unpaid resources gaining a lot of traction.

E-learning is widely controversial: is it beneficial, or not? According to some people, e-learning can turn education into a 'shared endeavour' [1], rather than just a teacher-to-student pipeline. Along with this, with the increased interaction and flexibility that online learning brings, many studies have found that retention of information is increased.

However, there are multiple arguments against e-learning as well. E-learning requires self-discipline, which many students lack—it is easy to change the tab on your computer to a video game or YouTube video during an online lecture. Along with this, it could also lead to social isolation [2].

The swift move to e-learning has excluded millions of students. According to a survey conducted by Child Fund, an NGO based in Delhi, 64% of rural Indian students might be forced to drop out due to a lack of external tutoring and/or assistance with acquiring technology [3].

Regardless of its effects, e-learning has created a place for itself in the lives of millions of students. In India, e-learning is elusive: only 24% of Indian households have access to the internet [4]. During times like the pandemic, e-learning is imperative. Not to mention, with the world moving quickly towards a more online platform for everything—work, transactions, school—it is important that technology is introduced into students' lives from the beginning of their journey.

Lack of awareness also accounts for a lot of these statistics: there are certain online resources that are available to all, but that most people—especially from lower economic backgrounds—are not aware of. I will be exploring these resources, along with paid resources, in this paper.

This paper will outline the state of e-learning and education in India, with an emphasis on K-12 learning—both for students with required resources to learn online, and those who do not have such an access. I will also explore the advantages and disadvantages of making e-learning resources available to all students, and the socio-economic impacts surrounding a decision like this, if it ever is made.

# 2. A Summary of the E-Learning Market in India

This section focuses on the e-learning market with a matrix, and descriptions of its subsets following it. The subsets have been colour coded along with their descriptions.

## E-Learning Matrix

| **User groups/ audiences** | **K-12 audience access to appropriate devices and bandwidth** | | | **K-12 with limited resources** | | | **Higher Education** | | | |
|---|---|---|---|---|---|---|---|---|---|---|
| **Type of learning** | **Test Prep** | **Supplemental education** | **Additional career options + subject choices** | **Test Prep** | **Supplemental education** | **Additional career options and/or subject choices** | **Formal degree programs** | **Skills based training with formal certification** | **Skills training with industry certification** | **Additional career options and/or subject choices** |
| **Providers** | • Byju's<br>• SAT N Paper<br>• Princeton Review<br>• Tutela, Career Launcher<br>• Bansal's<br>• Aakash Institute | • CBSE<br>• NCERT<br>• Corporate platforms (IBM, Microsoft, Adobe…) | • University programs<br>• Coursera<br>• Masterclass<br>• Career fairs | Free or lower priced platforms:<br>• CBSE<br>• NCERT<br>• Khan Academy | Free or lower priced platforms:<br>• CBSE<br>• NCERT<br>• Khan Academy | • University programs<br>• Coursera<br>• Masterclass<br>• Career fairs | • Internal university programs<br>• Edx | • Internal universities and their online degrees<br>• Edx<br>• Mitx<br>• Coursera<br>• Udacity | Corporate platforms like:<br>• Skillsbuild<br>• NASSCOM<br>• Futureskills | A combination of all |
| **Current Scale (Present market size)** | 15 million | 3.7 million | | 15 million | 3.7 million | | 5.5 million | 3 million students | | 8.5 million |
| **Potential scale (Total Addressable Market)** | 250 million | 9.5 million | 250 million | 250 million | 9.5 million | 250 million | 37 million | 37 million | 37 million | 37 million |
| **Cost to user vis-a-vis traditional** | Similar | Lower | Lower | Higher (acquisition of devices) | Higher (acquisition of devices) | Lower | Lower | Lower | Lower | Lower |
| **Cost to provider vis-a-vis traditional** | Dramatically lower on a larger scale, otherwise slightly lower. | Higher | Lower | Lower | Lower | Lower | Lower | Lower | Lower | Lower |
| **Perceived benefits** | • Convenient and time saving (no commute)<br>• Access to global resources and educators | • Can be accessed without time or place constraints<br>• Is (mostly) free, in | • No time constraints (sessions are recorded)<br>• Students can complete at home | • Convenient and time saving (no commute)<br>• Access to global resources and educators<br>• Cheaper | • Can be accessed without time or place constraints<br>• Is (mostly) free | • No time constraints (sessions are recorded)<br>• Students can complete at home | • Cheaper<br>• Staying at home<br>• Work can be done on student's schedule<br>• Certification is the same as a real-life degree | • Builds skills in easily digestible courses<br>• Cheaper than attending a workshop for the skill<br>• Brings job opportunities to | • Builds skills in easily digestible courses<br>• Cheaper than workshops | • Would build knowledge in other avenues<br>• Certification would look good on resume |
| | • Instantaneous results on mock tests | many cases personalised<br>• Helps build study habits<br>• Enables knowledge retention | (no housing cost etc)<br>• Course credits (if any) enable earlier graduation | • Instantaneous results on mock tests | • In many cases personalised<br>• Helps build study habits<br>• Enables knowledge retention. | (no housing cost etc<br>• Course credits (if any) enable earlier graduation | | the table (certification)<br>• Assignments ensure proficiency | | • Can be completed from home |
| **Perceived challenges** | • Rigor (more distractions for students and educators)<br>• Focus of educator on group instead of individual student<br>• Less interactive so lower retention of information. | • Students can be easily distracted<br>• Lacks deadlines so courses are not completed<br>• Less interactive than real-time classes. | • Fewer discussions with professors<br>• No exposure to student life at campus<br>• No attention on individual student<br>• Work graded by software instead of teacher | • Automated so less attention on student<br>• Language barriers<br>• Less rigorous than in-person classes | • Students can be easily distracted<br>• Lacks deadlines so courses are not completed<br>• Less interactive than real-time classes. | • Fewer discussions with professors<br>• No exposure to student life at campus<br>• No attention on individual student<br>• Work graded by software instead of teacher | • No exposure to college life<br>• No bonding/discussions with professor<br>• No schedule could mean easier to drop out | • Individual course completion could be disadvantageous (for skills such as public speaking) | • No certification means that it cannot go on the resume<br>• Individual course completion disadvantageous | • Takes up time that could be used for coursework<br>• Lack of certification would not increase employability chances |
| **Strategies employed to mitigate challenges** | • One to one class<br>• Small group classes<br>• Quizzes | • Quizzes added to course<br>• Tests and assignments to be completed within *x* amount of time post completion of course. | • Emails can be sent to teachers<br>• Cap on number of students present<br>• Classes done on discussion-friendly platforms | • Quizzes added, courses using pictures rather than words<br>• Deadlines given for homework | • Quizzes added to courses<br>• Tests and assignments to be completed within *x* amount of time post completion of course. | • Emails can be sent to teachers<br>• Cap on number of students present<br>• Classes done on discussion-friendly platforms | • Real-time calls to simulate a discussion environment<br>• Deadline on when lectures should be watched. | • AI could be used to detect tone and give feedback for specific skills | • Automated badges with proficiency level given at the end of a course<br>• AI could be used | • Automated badges<br>• Could be an easier side course (not as time consuming) |
| User groups/ audiences | K-12 audience access to appropriate devices and bandwidth | | | K-12 with limited resources | | | Higher Education | | | |

## Kindergarten to Twelfth Grade (K-12)

For test preparation within the K-12 with resources sections, providers are quite versatile. Providers include online platforms (paid ones like Byju's or free ones like Khan Academy), along with test prep centers like Tutela Prep, an agency based out of Gurgaon, India, and Aakash Institute. As of mid-2021, most of these centers are running online classes on platforms like Zoom and Microsoft Teams. For test prep among K-12 without apt resources, the providers are limited, with free or subsidized platforms like CBSE, NCERT, and Khan Academy leading the way. For this market, lower priced test centers also take the front seat for tests like the JEE or NEET, although there are a vast number of other centres.. The current scale of test prep for both K-12 with and without resources is 15 million people [4], although this could scale up about 17 times to the total number of students in the system, which is 250 million [4].

As for online classes and their cost vis-à-vis the traditional, offline method of learning, for K-12 audiences with appropriate resources, the cost is similar, if not lower. This is due to device and bandwidth availability prior to online education becoming a necessity. Along with this, transportation costs are also cut down dramatically. For K-12 without apt resources, however, this changes, since many must acquire laptops or tablets to attend these classes, along with the often out-of-budget fees charged by a lot of test centers, even for online classes. With online platforms like Khan Academy, though, the price of learning online is lower, regardless of resource status.

There are also significant benefits to learning online. As mentioned, online learning can be cheaper, and time can be saved for commute. Along with this, online resources are available from across the world, making online education a better avenue in that aspect. Online education is also often cheaper, especially if it is on a paid-for online platform like Byju's (which also provides free courses) or Kumon. The removal of human educators and the addition of AI also creates faster grading systems and feedback, which can be useful.

That said, there are also caveats. There is a significant lack of rigor in automated online platforms, along with less focus on one student if classes are being conducted by test prep centers. There is also far more distraction at home, which can be detrimental. The lack of interaction can also create a significant lack of interest. If an automated platform is being used, language barriers can also cause inaccessibility. These problems can be solved with one-to-one classes and quizzes, though, which let the student know their progress and/or motivate them. To solve inaccessibility, courses can rely more on visuals, less on language, to ensure understanding despite linguistic differences.

As for supplemental education, the market is smaller. Supplemental education would be defined, in this case, as academic help that augments a student's learning experience. There are several platforms here: including the test prep centers I mentioned above, along with platforms like CBSE, NCERT, and corporate platforms (like IBM, Microsoft, or Adobe). The platforms are similar for both K-12 audience with and without access to appropriate resources, excluding paid tuition academies like Tutela prep or Byju's for K-12 without resources. As of now, the online audience for supplemental education is approximately 3.7 million students, which could potentially increase to 9.5 million students using online means to study. Alternately, it could also increase to 250 million schoolchildren, approximately 63 times its current size. For online supplemental education, the cost to the user is typically lower since most platforms offer it for free. However, the cost to the provider may be higher (i.e., they would have to build a platform that could handle courses).

The benefits also outweigh the challenges. Supplements can be accessed without any time or destination constraints, can be personalized, build individual study habits, and enable knowledge retention [6]. As mentioned earlier, they are also dramatically cheaper, both in terms of transportation and in terms of tuition. The challenges can easily be mitigated. One challenge would be that distractions are more readily available, and for this, self-grading quizzes could be added to courses. Another concern may be that it is easy for the student to forget about their work since there are no deadlines. I have faced this myself. To combat this, deadlines could be added to courses. If they are not fulfilled, the course could time out, and the student would have to start again, making the cons of skipping homework outweigh the pros.

The last subset in the matrix is “Additional career options and/or subject choices”. This refers to summer programs from universities, courses on platforms like Masterclass or Coursera, or career counselling. While all of these are available to students with access to appropriate resources, only free programs from websites like Coursera are available to students from limited resource backgrounds. While I could not find an exact number for the current market of additional career options and subject choices, the total addressable market would be 250 million schoolchildren. The cost of conducting paid additional career option courses (such as online college fairs or subject-wise orientations) would be lower in price for both the provider and user, since offline, conducting these courses also would have included housing and maintenance. Platforms like Masterclass and Coursera, which rely entirely on online revenue and audiences, will not see any change in cost. The benefits of attending these courses online would be that there are no time constraints, they are cheaper (no housing costs, as mentioned earlier), may carry course credits that could enable earlier graduation. Caveats in this sector would be that there are few discussions with professors, which could create a flawed idea of the interaction required at a higher-education level, a lack of individual attention, and work graded by a software would lack constructive feedback. To mitigate these, students could send emails to teachers, the professors could have a cap on the number of students allowed in their courses and conduct classes on platforms like Zoom, where the class could be open to discussions.

## Higher Education

Higher education is also a significant player in the e-learning market. The current market for online formal degree programs is approximately one-seventh (5.5 million students) of the total addressable market, which is 37 million students enrolled in higher education [7]. The cost of conducting these formal degree programs online is lower for both provider and user for the same reasons that career counselling is—there is no cost for housing. Some perceived benefits of receiving an online degree are that it is cheaper, can be done on any schedule, and the degree is the same, with no distinction as to whether the student attended online or offline classes. Some disadvantages would be that the student would not be exposed to college life and would have no discussions with their peers or their professor. The lack of schedule could also make it easier for the student to drop out of school. A strategy to mitigate that would be to have real-time lectures which simulate a discussion environment and required hours to speak to fellow online learners and the professor. There could also be a deadline for when a lecture and the homework given with it would have to be completed.

Another player in online education for higher education students is skill-based training. Skill based training comes in two forms: with formal certification, and without. As of now, approximately 3 million students are enrolled in vocational training. This could be expanded to 37 million higher education students. The cost to the user, as compared to offline training, would be lower with similar results for the provider. An advantage to both types of skill-based training would be that they build skills in easily digestible courses, are cheaper than attending a workshop on the same topic, and assignments ensure proficiency. A plus in formally certified skill-based training is that it can also bring job opportunities to the student. However, disbenefits are that individual course completion is disadvantageous, especially for lingual courses where speaking is necessary. For skill-based training that does not provide certification, another limitation is that the lack of certification means that the skill cannot be shown as certified on a resume, which means that it is not advantageous for finding employment. Strategies to solve these issues are to use AI tone and inflection detectors (such as the ones on online platforms like Duolingo) which can give feedback on oratory skills. For training without certification, automated badges of proficiency could be given at the end of a course.

The last column of my matrix is about career counselling or subject choices for further education for students in higher education. Platforms for this would include all the ones I have mentioned in this part of my research paper. Currently, the market is of 8.5 million students in this avenue. I reached this number by totaling the number of skill-based training program students and the number of students enrolled in online formal degree programs. The total addressable market would be 37 million students, and the cost to both parties would be lowered. The benefits of this are that it would build knowledge about other subjects that are not pertinent to the student's major, the certification would make an addition to the student's resume, and it can be completed from home. A challenge would be that these would take up time that could be used for schoolwork, and if the courses do not provide certification, would not increase employability. However, these can be mitigated through automated badges, and decreasing the difficulty of the course to make it less time consuming.

# 3. Research: Data and Methodology

In this paper, I investigated e-learning providers and the effects of e-learning on students and educators, especially since the start of the COVID-19 pandemic. I collected primary data through a survey circulated within my circle of students and teachers. I also interviewed parents to ask about the changes in fee structure and expenditure on education in general.

In the survey, participants primarily answered multiple choice questions, with a few number-scale questions as well. I selected my audience through cluster sampling—they were from the K-12 with resources section—and conducted the survey on Google Forms. In total, there were 142 respondents, from an audience of 800 students (a response rate of 18%).

For the interview process, I selected participants from different socio-economic groups. The first participant was a parent who sent his child to an international school in Gurgaon, and the second was a parent who sent his children to a boarding school in rural India. The interviews were taken over the phone, approximately ten minutes each, and I recorded them by taking notes. The interviews were informal, but they were semi-structured since I did have a few questions in mind.

My secondary data was collected through online resources, such as news articles. This was primarily for Part 2 of this research paper, where I explain the providers, options, and challenges for each type of education if taken online—divided by K-12 with and without resources, with higher education undivided.

# 4. Results

With my research, I hoped to reach a conclusion on whether providing widespread access to all groups would be beneficiary or harmful. Before that, however, I conducted a survey based on the experiences with online learning of different socio-economic groups. This was to understand the impacts of online learning and teaching on students and teachers to reach an accurate conclusion on whether online education is beneficial.

This was a multi-pronged survey: I asked students and teachers from higher-income backgrounds what their experiences with online learning were, but I also needed to question people from lower-income backgrounds how they coped with online learning.

My survey with people from higher-income backgrounds yielded interesting results. Out of approximately 800 people asked, 142 people responded to my survey (a turnout rate of 18%). Out of this number, 103 were students, and 39 were teachers.

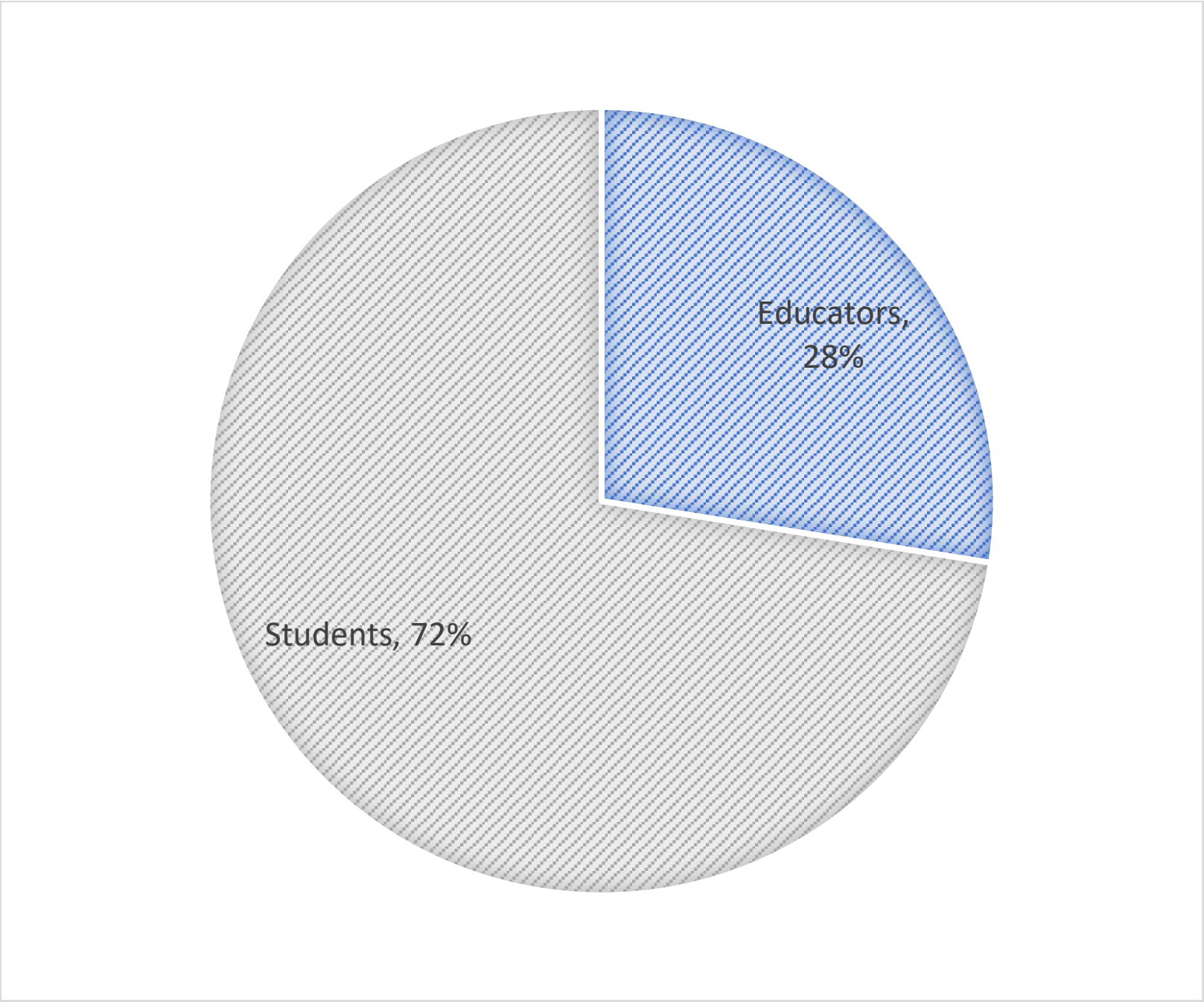

Everyone had access to technology and appropriate internet bandwidth prior to the COVID-19 pandemic, which was when most of them started learning/teaching online. 63.2% of the teachers started teaching online post the beginning of the pandemic (March 2020), while 69.9% of the students surveyed started learning online at the same time. Among teachers, a majority used Microsoft Teams (61%), with students (83%) having a similar outcome.  Other popular platforms included Google Meet (29% for teachers, 5.8% for students) and Zoom (5.3% for teachers, 12% for students).

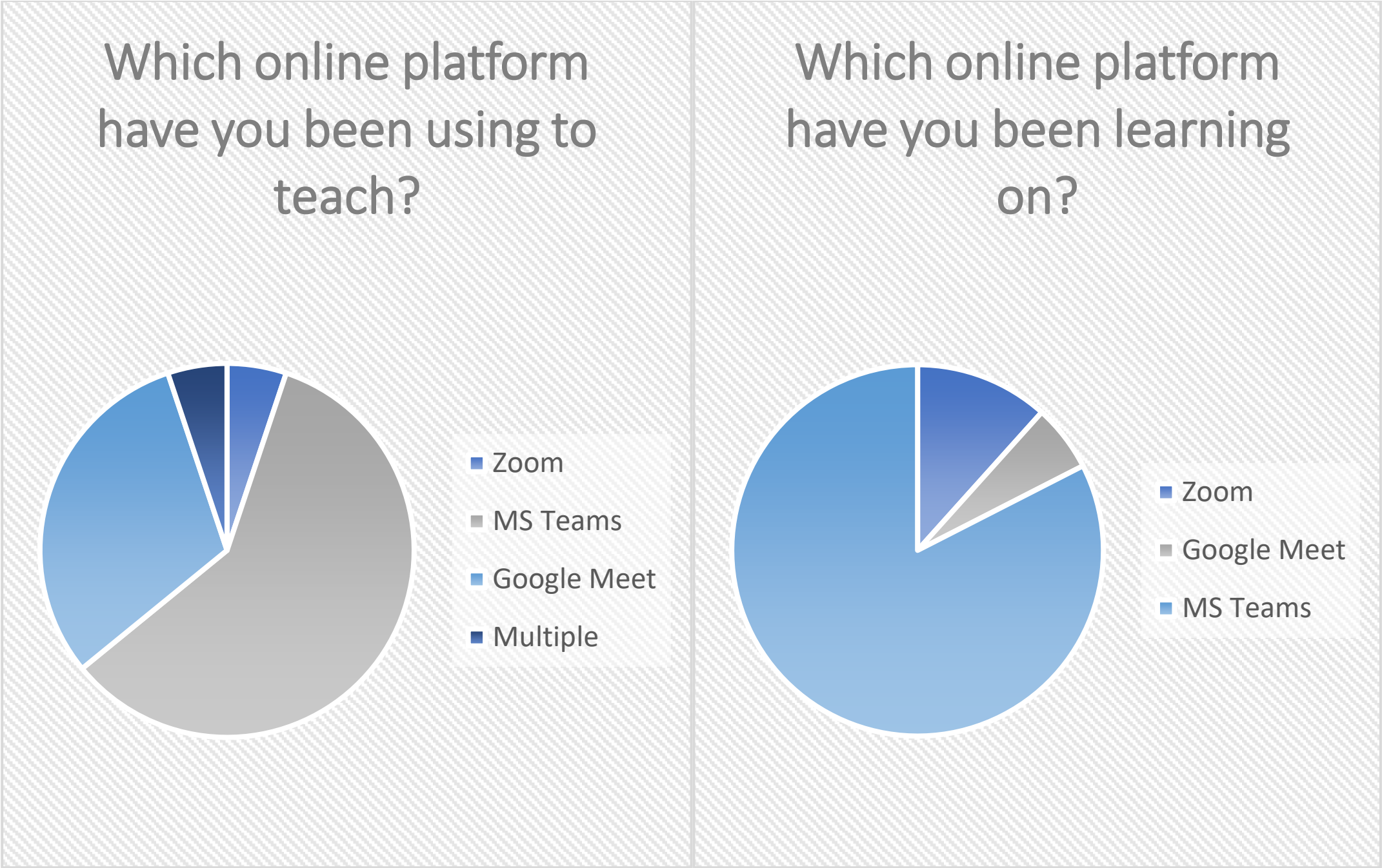

Teachers on the left, students on the right.

However, despite using Teams the most, students showed more comfort using Zoom (45%) when asked which platform they were most comfortable with. That said, Teams was a close second (38.8%). Google Meet was also popular, with 16% of students voting on it. Teachers, on the other hand, were split between Google Meet and Teams when asked the same question, with 37% voting on both, forming a combined percentage of 74%. Zoom was in second place, with 21% of teachers voting on it.

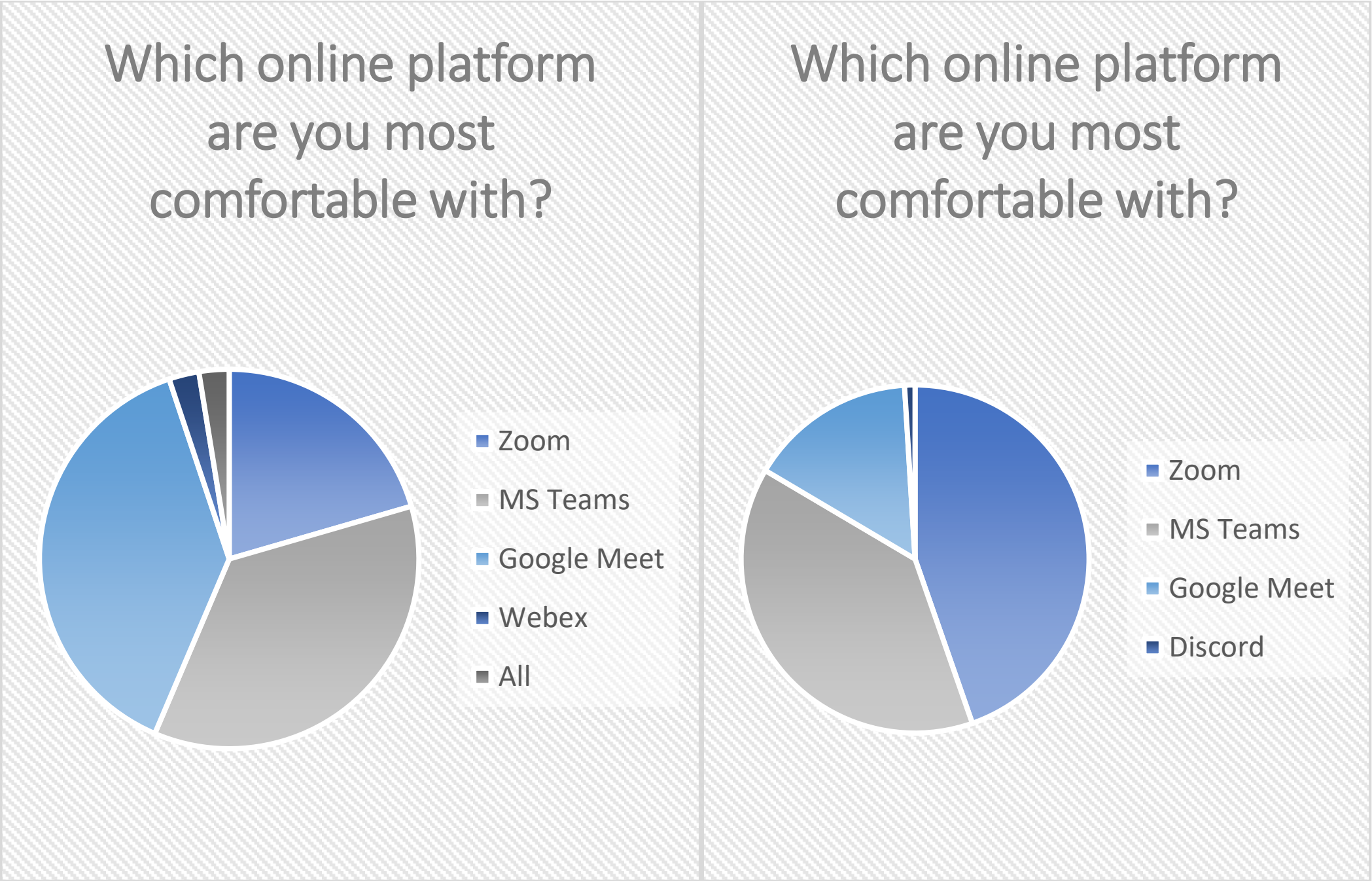

Teachers on the left, students on the right.

The next question I asked both groups was whether their transition to online education from offline education was easy. This was on a scale of 1-10, 1 defined as ‘Easy’ and 10 as ‘Difficult’. The mean for this among both students and teachers was 5.7 out of 10, showing a medium level of difficulty overall. However, the mode was the 7-10 range, showing that a majority had a difficult time starting on e-learning platforms. The results of this question reveal that both students and teachers faced an equal amount of difficulty in beginning online learning/teaching, but I also wanted to investigate how well both groups had adapted to online learning.

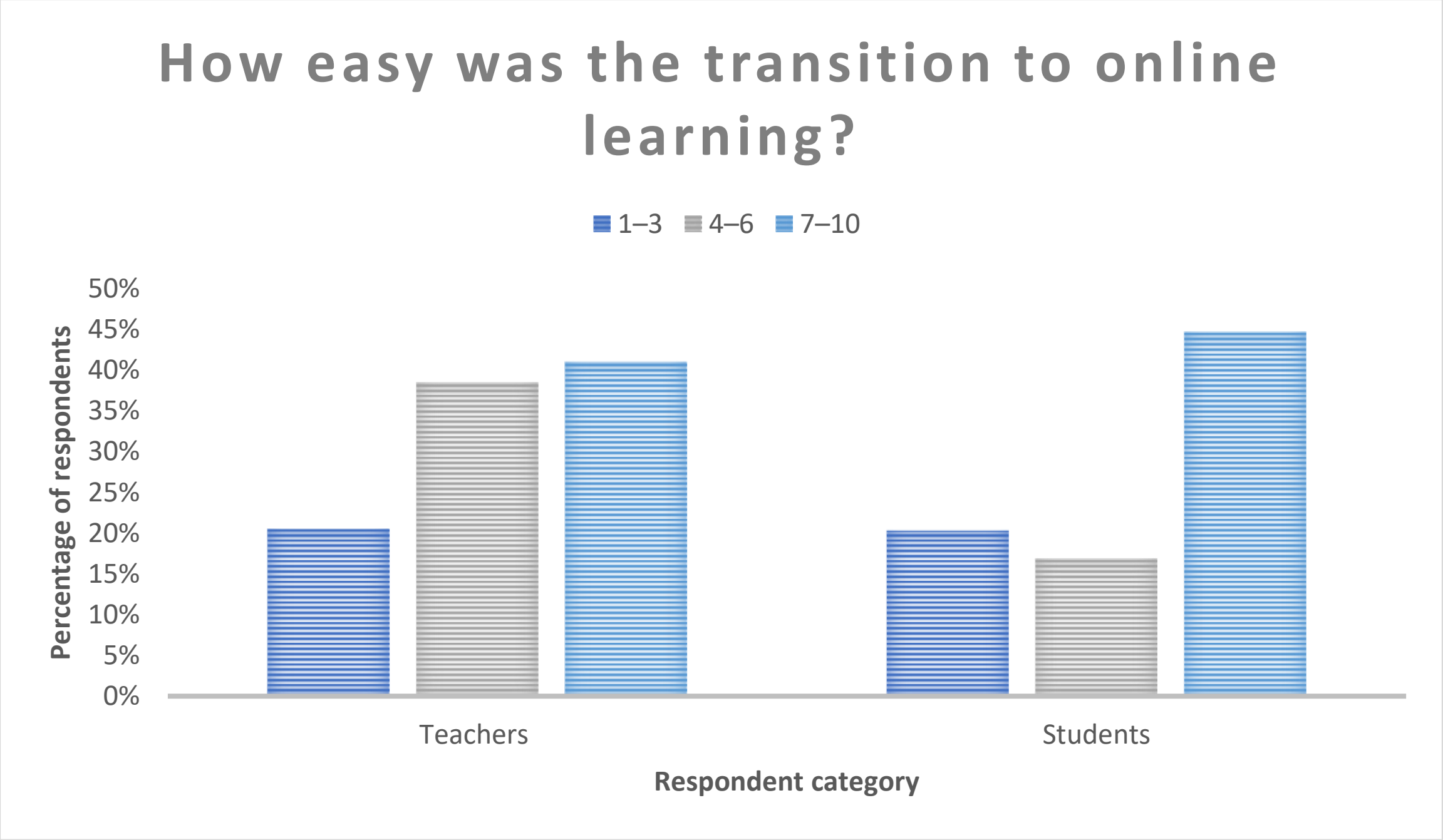

Therefore, my next question was how comfortable they were with learning/teaching online currently (as of May 2021). This was another scale question, with 1 defined as ‘Very uncomfortable’ and 10 defined as ‘Very comfortable’. While a clear majority of teachers said they were very comfortable, students were more distributed. For students, the mean was 5.9 out of 10, with teachers feeling more comfort at a mean of 7.2 out of 10. This was a surprise to me, since my hypothesis was that teachers would face more discomfort at the thought of online learning as compared to students, based on my own experience with teachers.

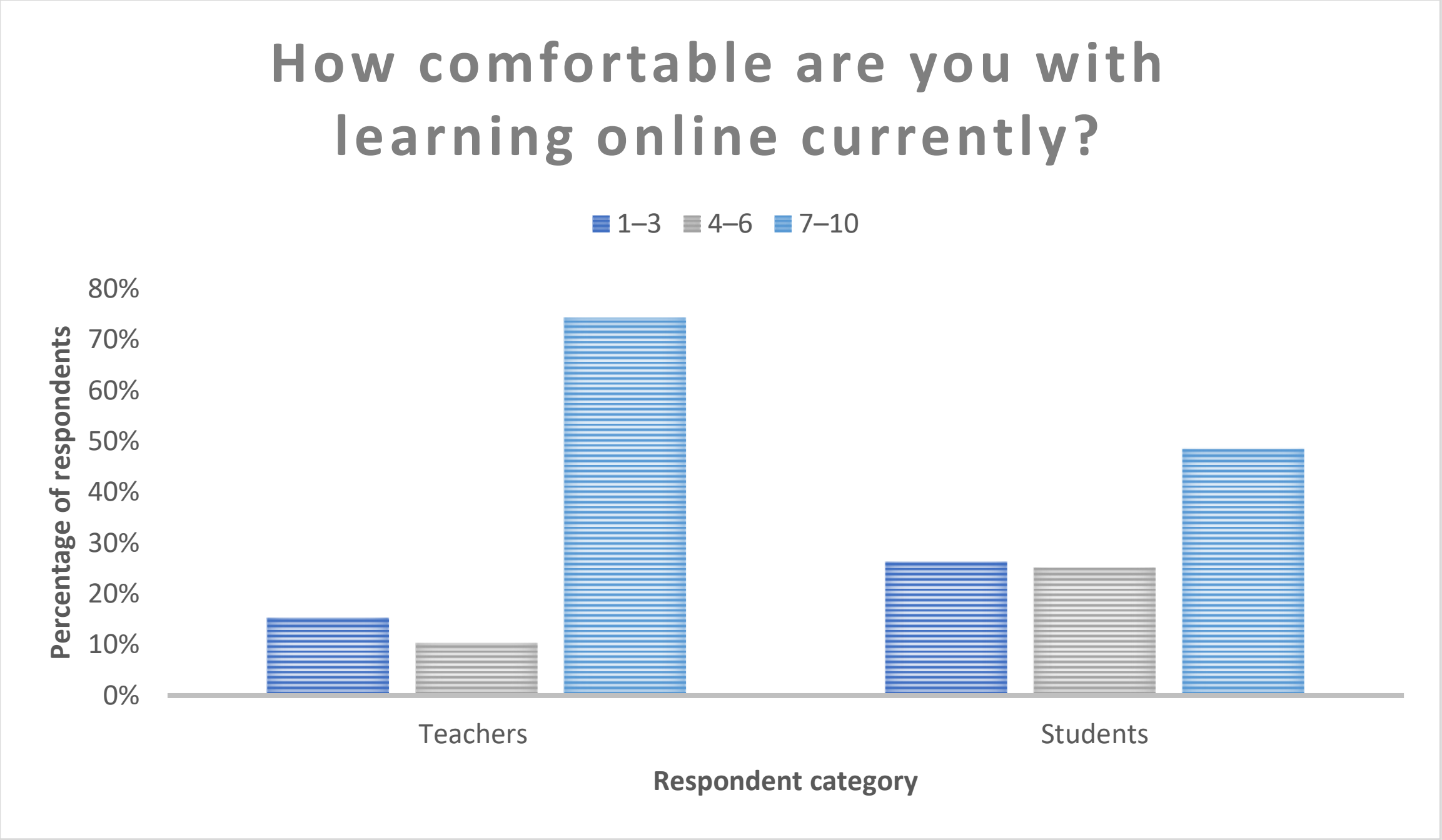

I also asked my respondents whether their learning/teaching style has changed, compared to how it used to be before COVID-19. The question was on a scale of 1-10, with one being ‘Not much’ and 10 being ‘A lot’. Most students (32) said that it has changed a lot, stating 10 on the scale, while 10 teachers said that it had changed by 8 on the scale. The mean for teachers was 6.9/10, while for students it was 7.8/10.

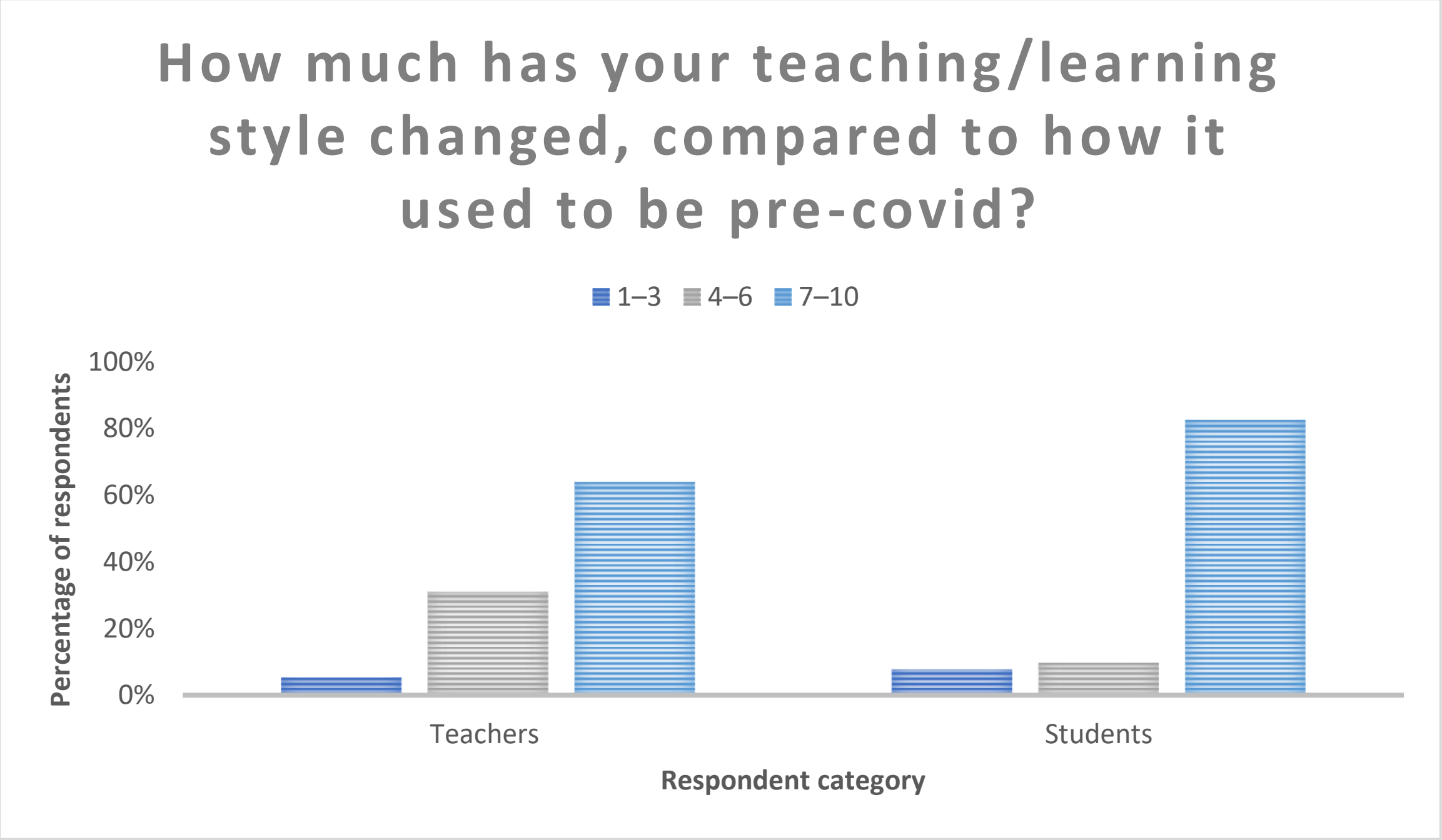

My next question was solely for teachers: “What method do you prefer to teach with online, and has this changed compared to before?” Here, I received varied answers. A vast majority of teachers agreed that their teaching style *had* changed due to teaching online. Most said that they had to use more interactive tools to keep their students captivated. Many preferred a discussion format of teaching in offline platforms but had to switch their styles up due to online platforms making it harder for these types of classes to occur. A lot of them also conducted flipped classes instead, where students would create presentations on the topic at hand and the teachers would provide feedback and any extra information needed.

The question that followed asked teachers and students which online resources they used. Most teachers said they used free online resources, like Khan Academy, also naming CBSE, NCRT and NIOS as suppliers of supplements. Along with this, many also talked about YouTube, paid e-resources like Lit Charts, and documents on the internet. Among students, responses were similar, except for online test sites like Papa Cambridge also figuring on the list. Among teachers, 92% said they would continue to use these resources after e-learning is no longer the sole method for online education. 87% of students had the same response.

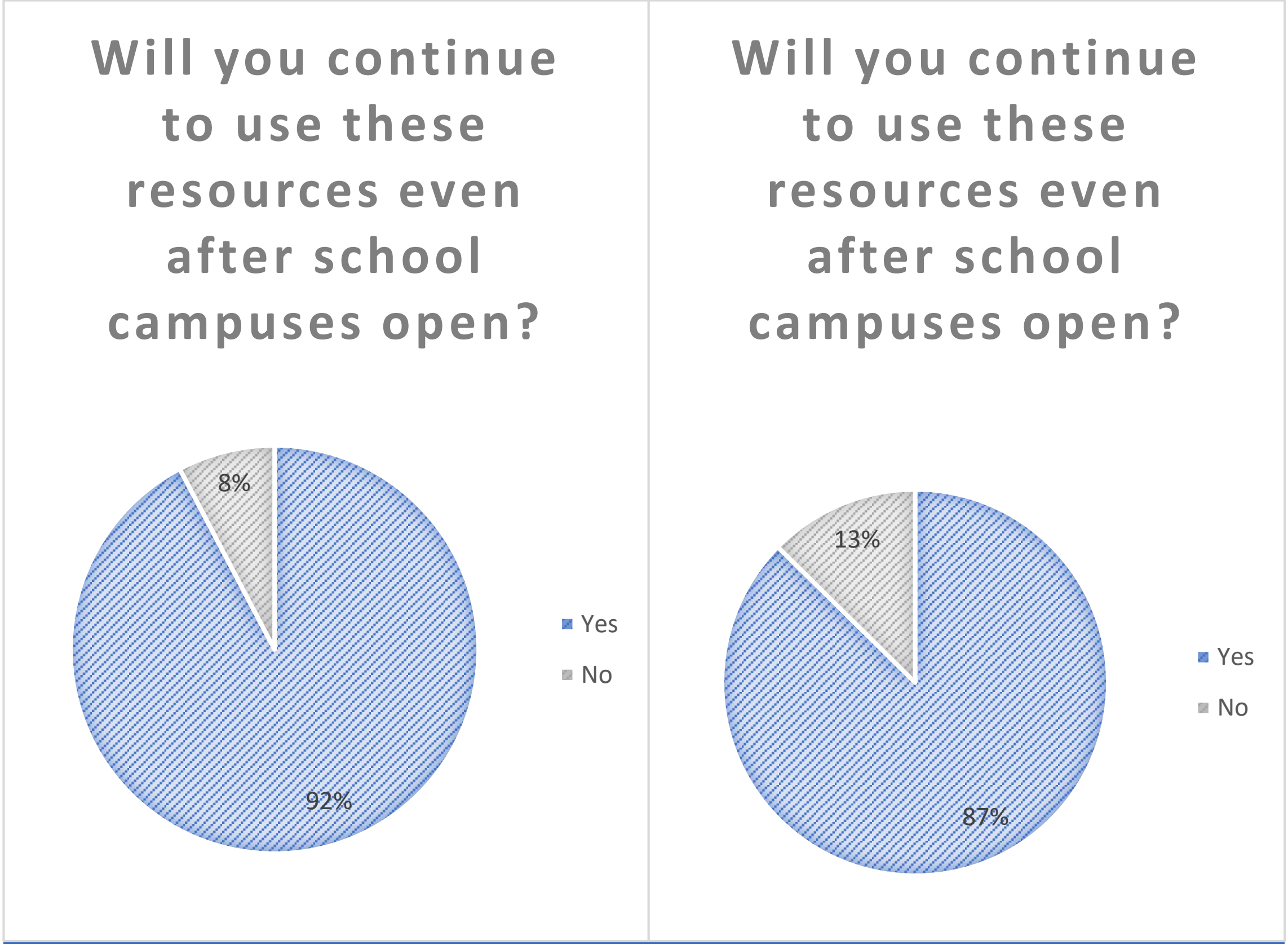

Teachers on the left, students on the right.

When asked whether they thought their students had benefited from online learning, 92.1% of teachers agreed that their students had benefitted to some extent. On the other hand, 54.4% of students believed that they would have had a better experience learning at school. That said, 27.2% of students also believed they had benefitted to some extent.

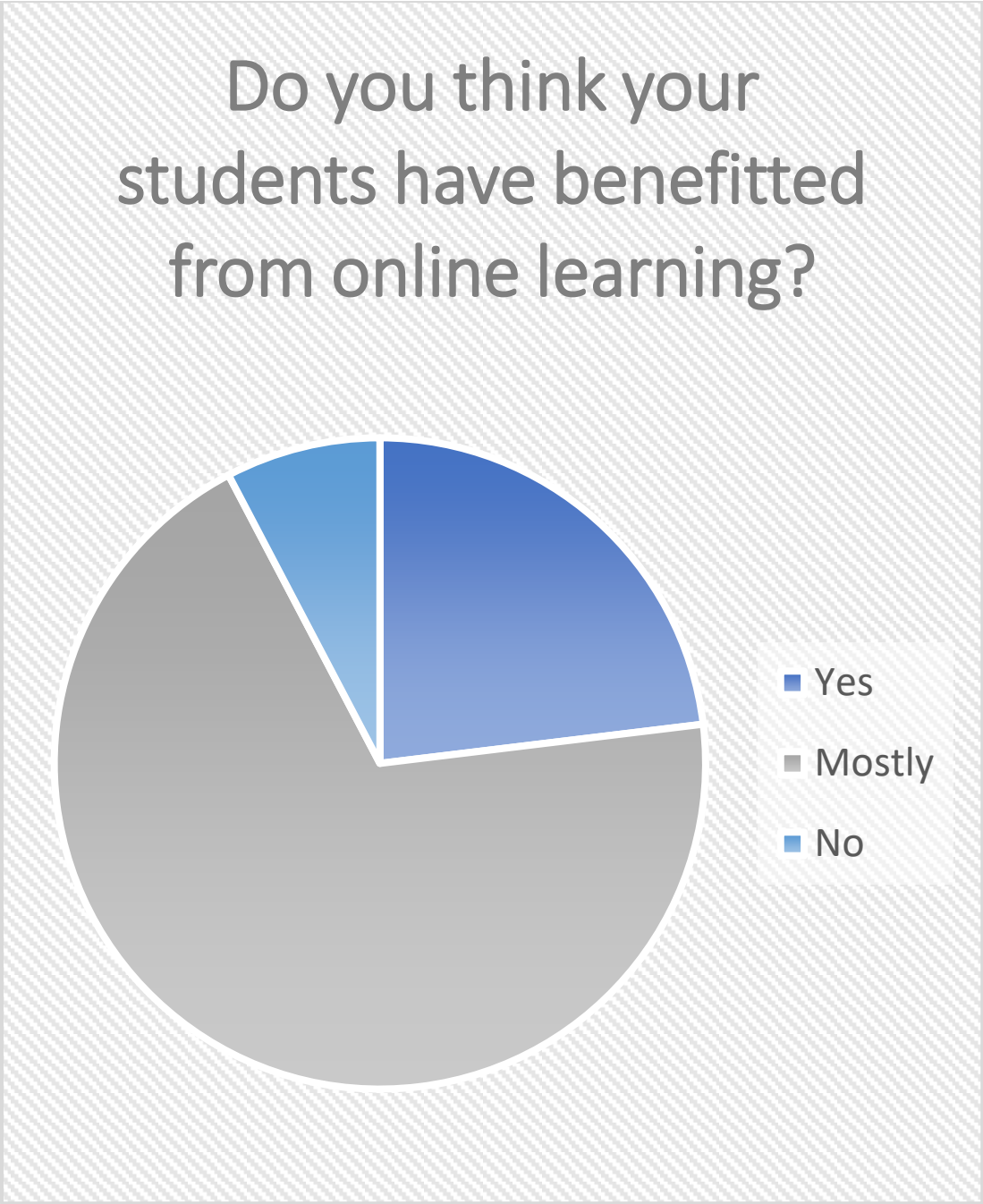

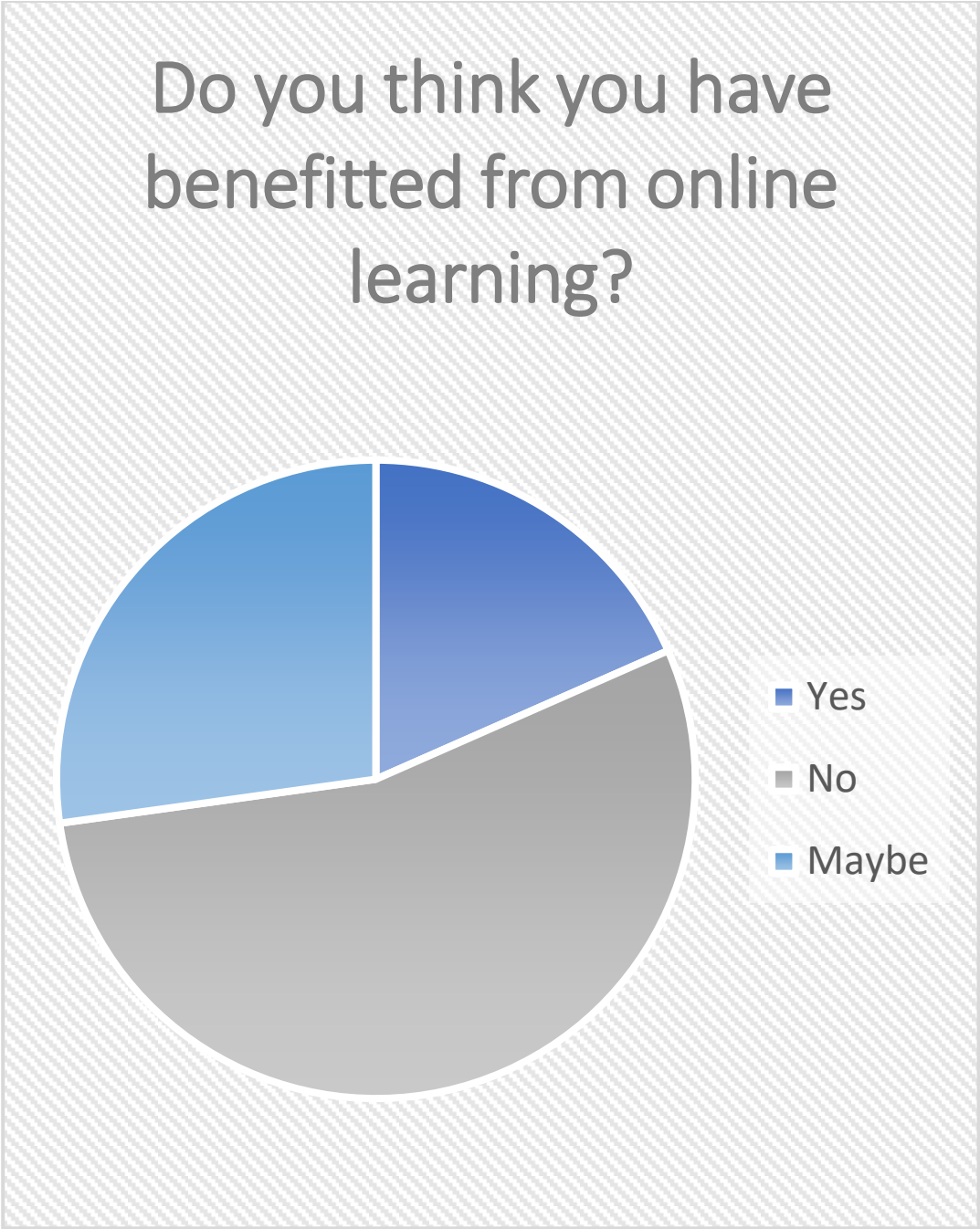

Teachers on the left, students on the right.

For my last question, I asked whether my audience would prefer to learn online, offline, or a mix of both if e-learning was not a necessity. 62.1% of students said that a mix would be best, with much of this number (82.7%) choosing a 50%, 50% online-to-offline ratio, or lower for online.  A larger proportion of teachers answered 'A mix' on the same question (81.6%), and of this number, 85.4% believed chose 50:50 online-to-offline or below for online.

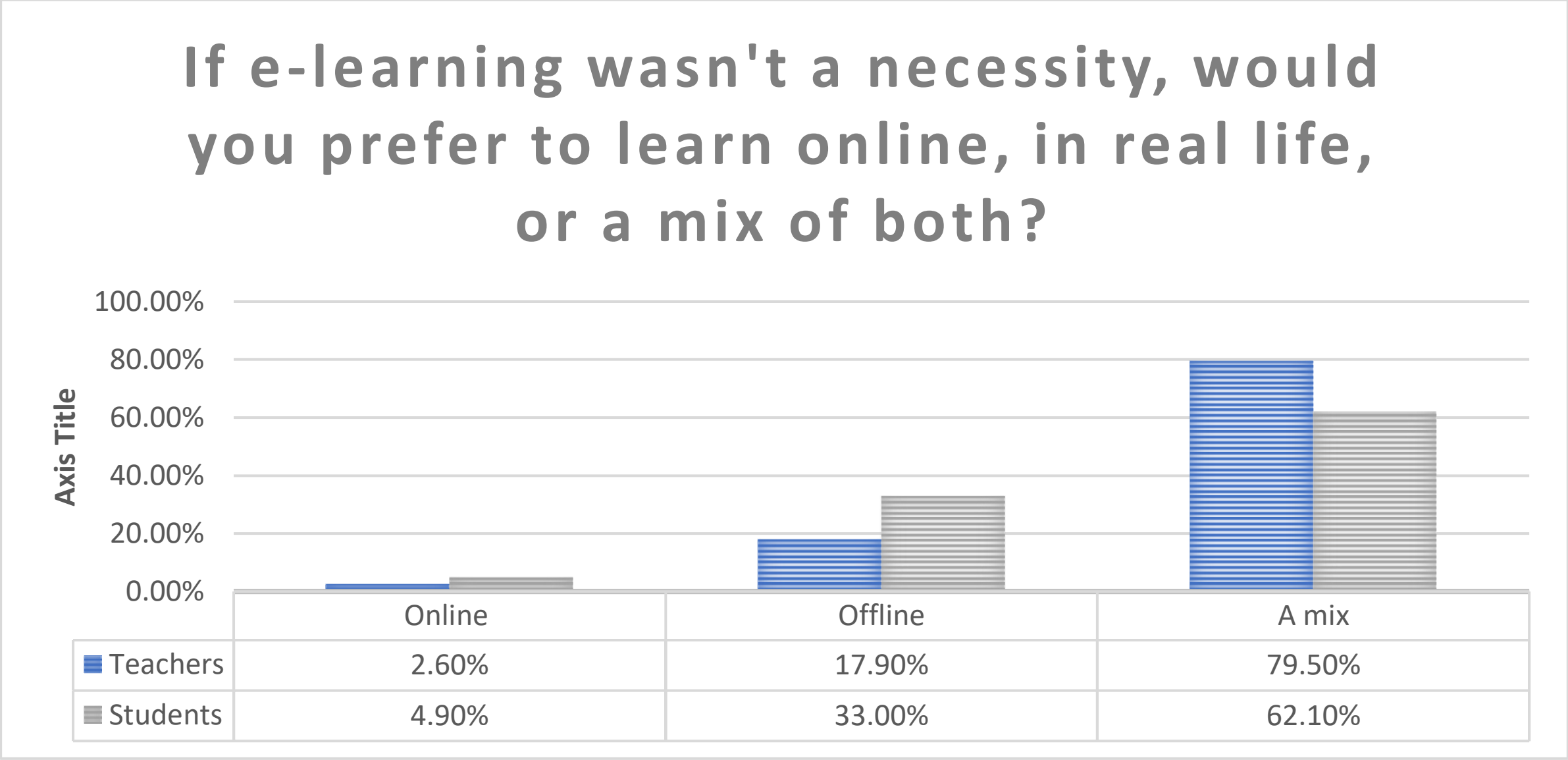

As for the differences in expenditure, I asked parents from an international school about how their spending has changed as compared to before the pandemic, during offline school. According to them, while lunch and bus fees had stopped, the school had started charging approximately 10,000 rupees more for education. So, overall, expenditure on e-learning was more than their expenditure on offline learning. That said, they did not have to spend money on purchasing a new device so online school could be conducted—they already had laptops available.

To see whether the outcome was similar for parents from lower-income backgrounds, I questioned a parent of two children who go to school in the city of Malda, in West Bengal. For them, school fees remained the same, and due to phones already being available, no devices had to be bought. The children also were not taking tuitions offline or online. However, based on field experience with an underserved school in Gurgaon, while fees did not change, many children were forced to buy their own phones when the COVID-19 pandemic first started. So, overall, for this economic background, expenditure on e-learning either stayed the same or increased slightly.

# 5. Providing Ubiquitous Access to All K-12 students

In this section of my paper, I will be providing options (and their challenges) for providing unanimous access to e-learning for all K-12 students.

## For ubiquitous access to hardware

- One option is for underserved schools to raise their fees, automatically gaining enough funds to provide students with devices and education on how to use online resources judiciously. However, a significant issue to this solution is that many K-12 students cannot afford the fee raise, which is why schools are left floundering.
- Another option would be to introduce donation drives for underprivileged schools, where old, good condition, hardware could be collected and reused by people who cannot afford it. A caveat to this would be that hardware is long-term, which means that donation drives would only be conducted once a year. Along with this, quality control would be difficult, and people may want to sell their old devices instead of donating them. However, this could easily be solved by placing volunteers, who could switch on the device and check if it is working.
- An option I find particularly interesting is large companies using their Corporate Social Responsibility funds to provide ubiquitous access to students. There could be potential for a middleman organisation for this, which could organise and obtain the funds.

## For awareness regarding e-learning

I believe that this is a more pertinent issue. Many children have phones through which they can obtain access to online platforms but lack awareness about e-learning platforms that could work better compared to expensive tutors without the correct qualifications. For this issue, the solutions are:

- Conducting workshops with various schools. This is a good option, especially since it is manageable. Instructional workshops could help students navigate the internet, and the various resources that are available to them.
- Educating parents would also be useful since many children do not choose their schools or their tutors. Parents could be informed of the various options regarding tuitions, and then they could, potentially, come to an informed decision.
- Subsidising online tuitions. I believe that this would be especially useful, since online subscriptions are often scoffed at, because e-learning is seen as inferior to offline learning. Subsidisation would encourage more traffic to these platforms—benefitting both provider and consumer.

## 6. Conclusions

In conclusion, the largest hurdles standing in the way of ubiquitous access to e-learning is a lack of infrastructure for e-learning to occur, and post that, a lack of awareness of the options available.

Based on experience—and the outcome of several studies—many students will be left behind during the pandemic because of the lack of infrastructure being awarded to them. The UN has estimated that, globally, more than 24 million school children are at-risk of dropping out due to the pandemic [9]. It is essential that these students are given the infrastructure to continue learning.

As discussed in Part 2 of this paper, there are many options available for students who lack the resources to pay for online education, or whose schools cannot afford a complete shift to online education from offline education. A lot of private coaches—hired by students regardless of socio-economic status—come without rating, promising training for various exams—the JEE, the NEET, the CAT, the SAT, the ACT being some—but delivering sub-par teaching. There are various online resources available that can replace private tutors, but a significant limitation for these platforms is the lack of awareness surrounding them. For many, it is more convenient to be able to visit a tutor close by rather than trusting an online platform due to wide-held beliefs about the trustworthiness of the internet. Therefore, it is important that students are not only given hardware to attend school-wise online education, but also awareness on the various supplementary education options.

My survey, and studies like it, prove that online education can be beneficial. Along with this, with the world moving towards a more online platform, it is imperative that access to online learning is made more inclusive, irrespective of socioeconomics.

# Annexure-I: Survey Questionnaire

## Section 1: Both Groups

1. Are you an educator or a student?
    - Educator (*form automatically skips to Section 2)*
    - Student (*form automatically skips to Section 3)8*

## Section 2: Teachers

2. When did you start teaching online?
    - Before the COVID-19 pandemic
    - After the COVID-19 pandemic started
3. Which online platform have you been using to teach?
    - Zoom
    - Microsoft Teams
    - Google Meet
    - Webex
    - Other
4. Which online platform are you most comfortable with?
    - Zoom
    - Microsoft Teams
    - Google Meet
    - Webex
    - Other
5. Was the transition to online education easy? Please indicate on a scale of 1 to 10 where 1 is "Easy" and 10 is "Difficult".
6. How comfortable are you with teaching online currently? Please indicate on a scale of 1 to 10 where 1 is "Very Uncomfortable" and 10 is "Very Comfortable".
7. How much has your teaching style changed, compared to how it used to be pre-COVID-19? Please indicate on a scale of 1 to 10 where 1 is "Not much" and 10 is "A Lot".
8. What method do you prefer to teach with online, and has this changed compared to before?

    (*Short answer question.)*

9. Which online educational resources do you use?
    - CBSE/NCERT/NIOS online resources
    - Free online resources like Khan Academy
    - Paid online resources (tutorial companies like Byju's, etc.)
    - All of the above
    - Other
10. Will you continue to use these resources even after school campuses open?

- Yes
- No

11. Do you think your students have benefitted from online learning?
    - Yes, most kids have benefitted
    - Yes, they have benefitted to some extent
    - No, it has been a waste of time
    - Other
12. If e-learning wasn't a necessity, would you prefer to teach online, in real life, or a mix of both?
    - Online
    - Offline
    - A mix
13. If you chose a mix of both, what ratio of online learning to offline learning would you prefer?
    *Choose the most appropriate answer. If none of the options work for you, please choose 'Other' and write down your preferred percentage*
    - 20% online, 80% classroom
    - 33% online, 67% classroom
    - 50% online, 50% classroom
    - 67% online, 33% classroom
    - 80% online, 20% classroom
    - Other

## Section 3: Students

14. When did you start learning online?
    - Before the COVID-19 pandemic
    - After the COVID-19 pandemic started
15. Which online platform have you been using to learn?
    - Zoom
    - Microsoft Teams
    - Google Meet
    - Webex
    - Other
16. Which online platform are you most comfortable with?
    - Zoom
    - Microsoft Teams
    - Google Meet
    - Webex
    - Other
17. Was the transition to online education easy? Please indicate on a scale of 1 to 10 where 1 is “Easy” and 10 is “Difficult”.
18. How comfortable are you with learning online currently? Please indicate on a scale of 1 to 10 where 1 is “Very Uncomfortable” and 10 is “Very Comfortable”.

19. How much has your learning style changed, compared to how it used to be pre-COVID-19? Please indicate on a scale of 1 to 10 where 1 is “Not much” and 10 is “A Lot”.
20. Which online educational resources do you use?
    - CBSE/NCERT/NIOS online resources
    - Free online resources like Khan Academy
    - Paid online resources (tutorial companies like Byju’s, etc.)
    - All of the above
    - Other
21. Will you continue to use these resources even after school campuses open?
    - Yes
    - No
22. Do you think you have benefitted from online learning?
    - Yes
    - Maybe
    - No, it would be better if I learnt at school
23. If e-learning wasn't a necessity, would you prefer to learn online, in real life, or a mix of both?
    - Online
    - Offline
    - A mix
24. If you chose a mix of both, what ratio of online learning to offline learning would you prefer?
    *Choose the most appropriate answer. If none of the options work for you, please choose 'Other' and write down your preferred percentage*
    - 20% online, 80% classroom
    - 33% online, 67% classroom
    - 50% online, 50% classroom
    - 67% online, 33% classroom
    - 80% online, 20% classroom
    - Other

## Section 4: Both Groups

25. How would you rate this survey? Please indicate on a scale of 1 to 10 where 1 is “Very Bad” and 10 is “Excellent”.

# Annexure-2: Survey Results

## Section 1: Both Groups

1. Are you an educator or a student?

   [Answers in the results tab.]

## Section 2: Teachers

2. When did you start teaching online?

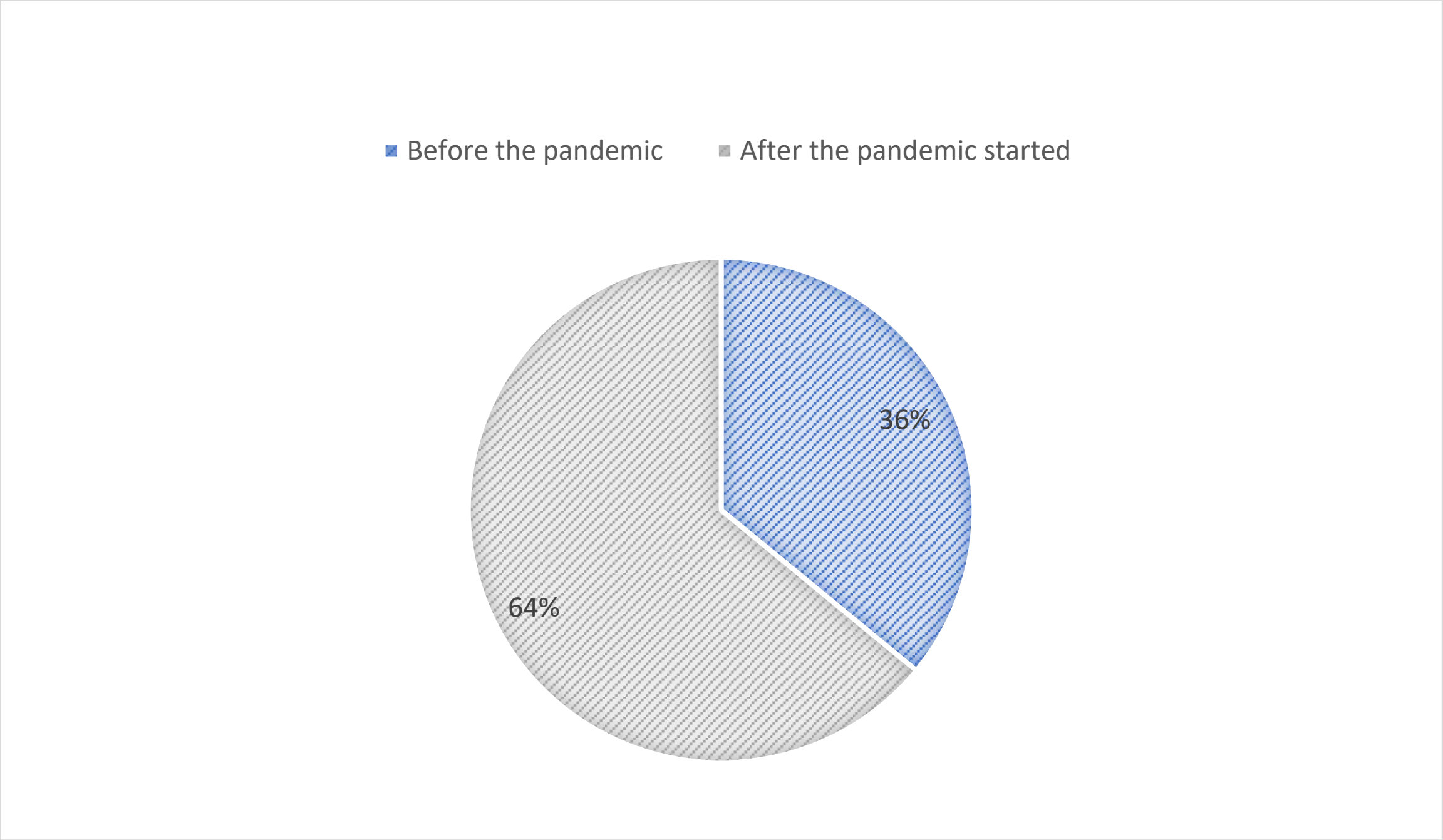

3. Which online platform have you been using to teach?

   [Answers in the results tab.]

4. Which online platform are you most comfortable with?

   [Answers in the results tab.]

5. Was the transition to online education easy?

   [Collated answers in the results tab.]

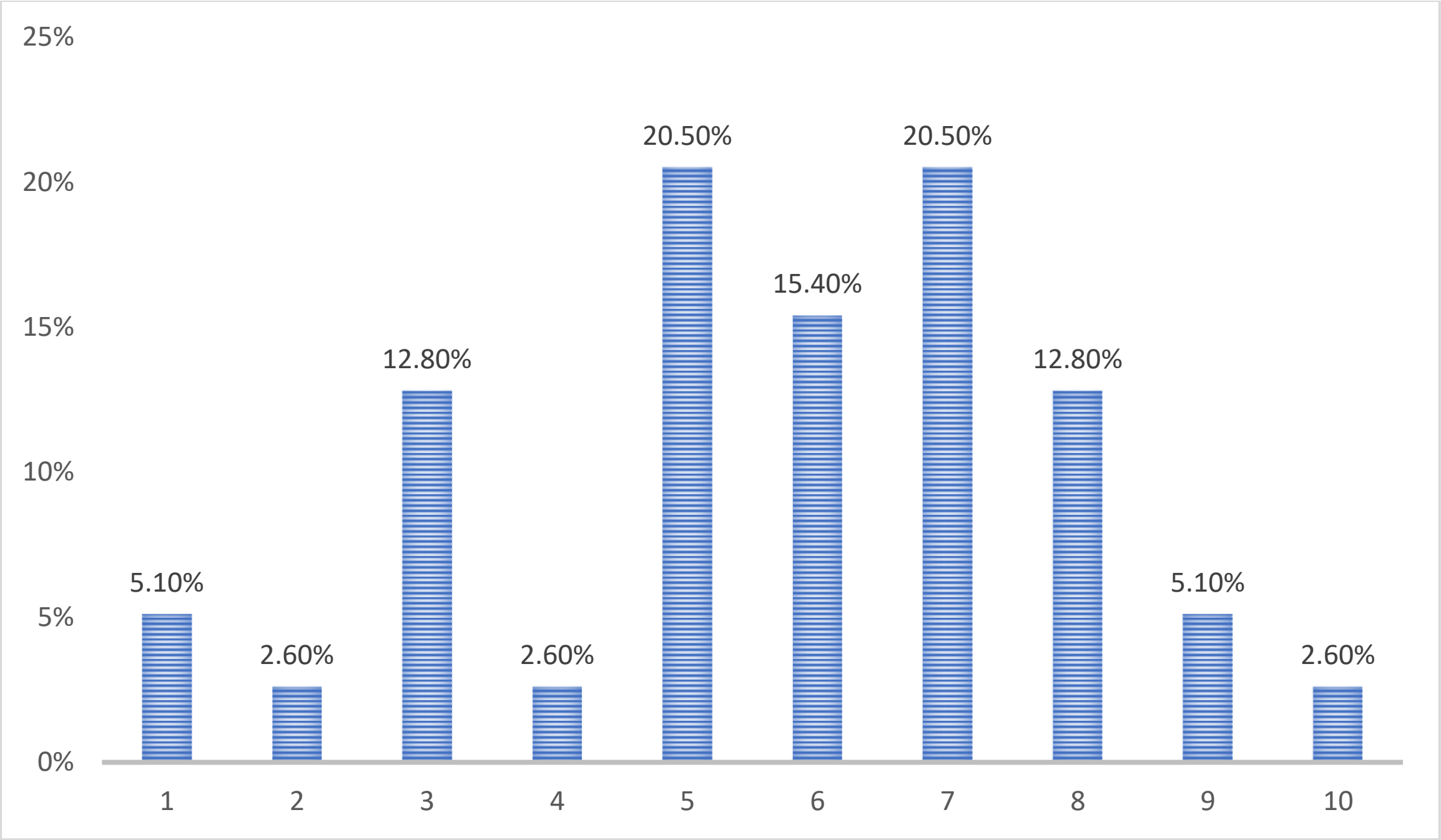

6. How comfortable are you with teaching online currently?

   [Collated answers in the results tab.]

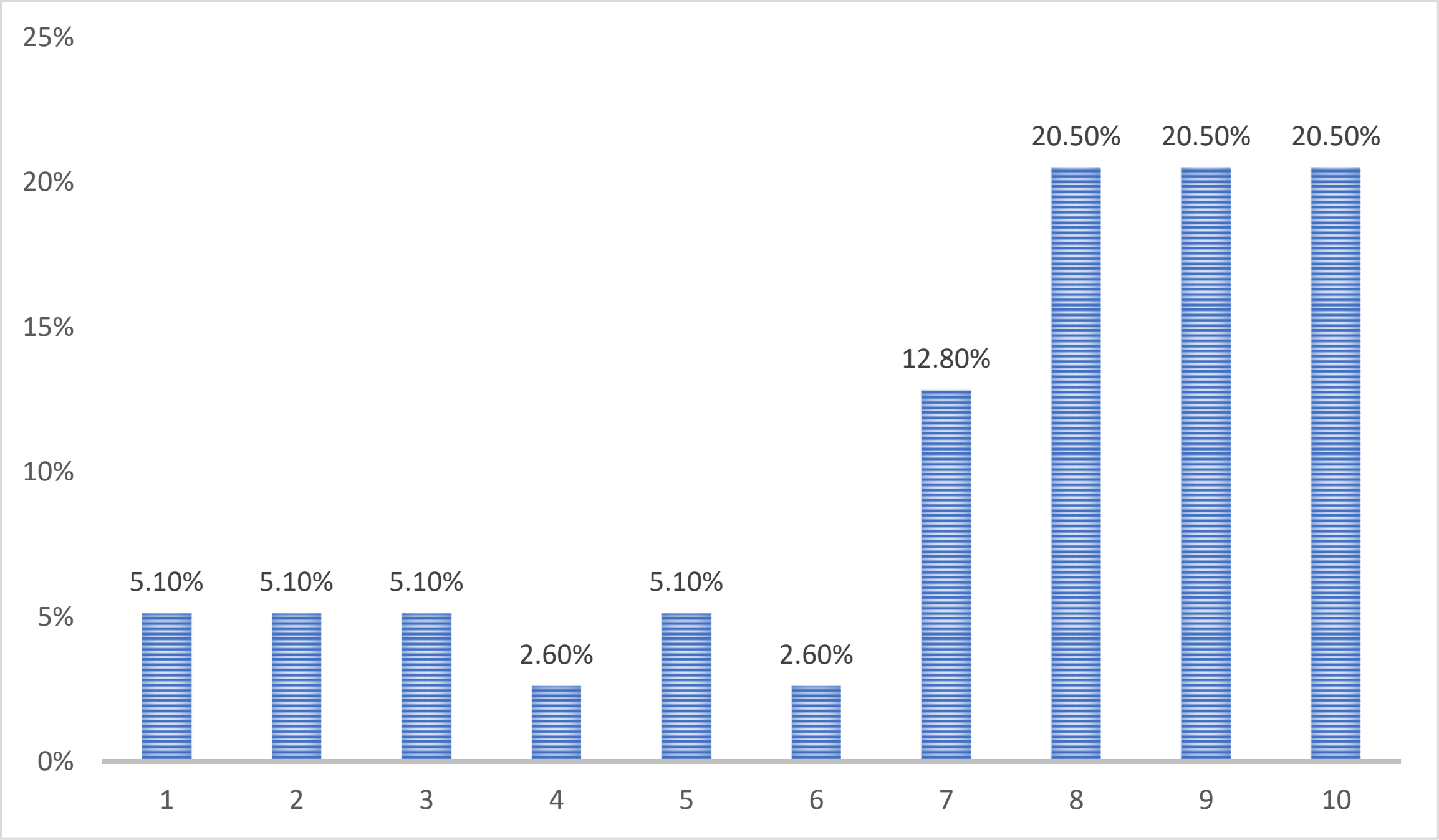

7. How much has your teaching style changed, compared to how it used to be pre-COVID-19?

   [Collated answers in the results tab.]

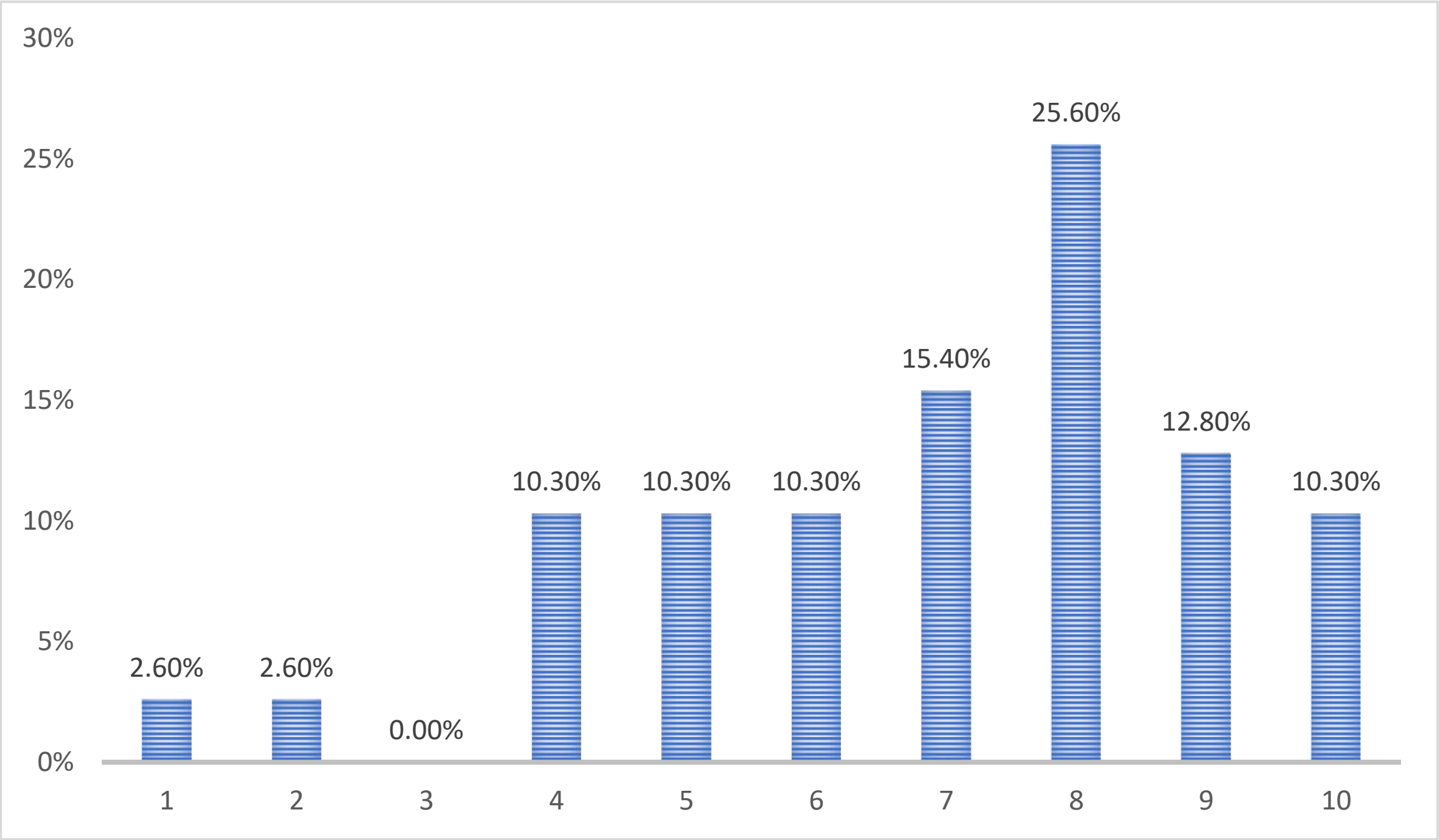

8. What method do you prefer to teach with online, and has this changed compared to before?

   *Short answer question, summarized:*

More visual (to keep students engaged) as compared to interactive earlier.

More project work and research.

More child-centric—flipped learning model.

Same, using an online blackboard instead.

More mentoring than teaching.

9. Which online educational resources do you use?

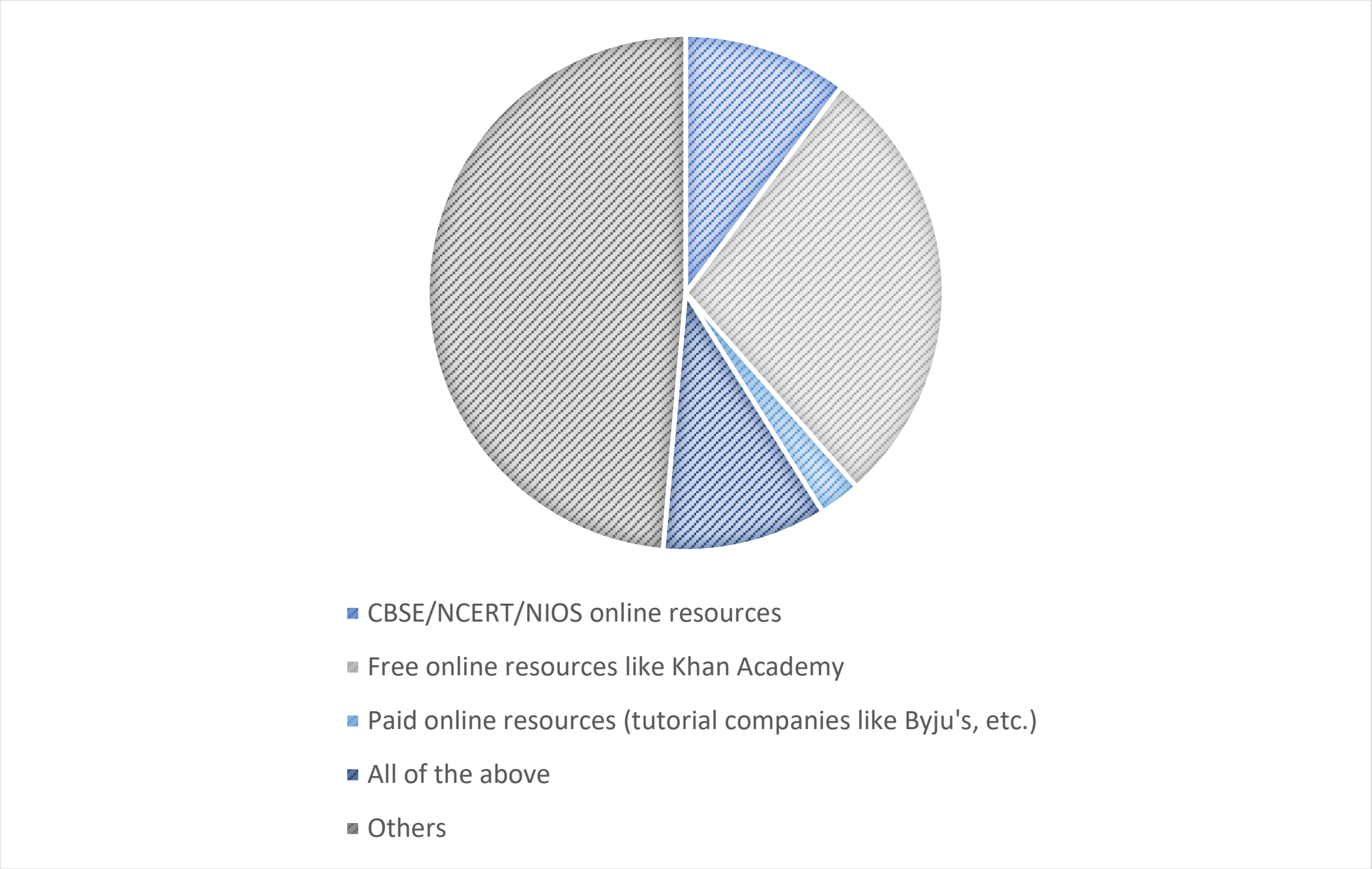

## Others

### Flashcards

- Quizlet

### Exam Archives

- Papa Cambridge
- Save My Exams
- Xtreme Papers
- IGCSE Teacher Resources

### Videos

- BBC Bitesize
- YouTube
- SkillsBuild

### Conceptual

- LitCharts
- Khan Academy

### Miscellaneous

- Personal tuitions

10. Will you continue to use these resources even after school campuses open?

    [Answers in the results tab.]

11. Do you think your students have benefitted from online learning?

    [Answers in the results tab.]

12. If e-learning was not a necessity, would you prefer to teach online, in real life, or a mix of both?

    [Answers in the results tab.]

13. If you chose a mix of both, what ratio of online learning to offline learning would you prefer?

    [Answers in the results tab.]

## Section 3: Students

14. When did you start learning online?

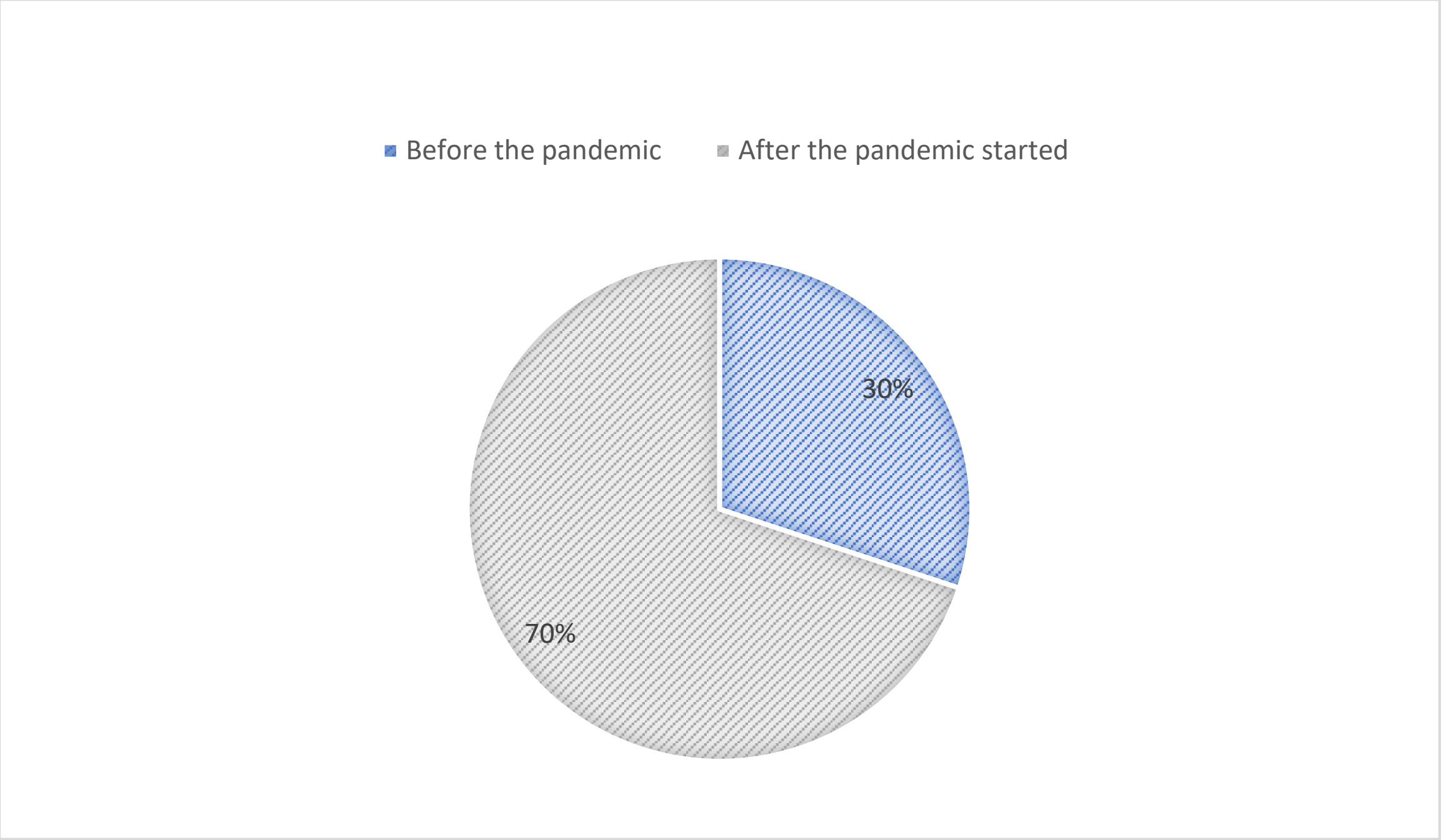

15. Which online platform have you been using to learn?

    [Answers in the results tab.]

16. Which online platform are you most comfortable with?
    [Answers in the results tab.]

17. Was the transition to online education easy? Please indicate on a scale of 1 to 10 where 1 is "Easy" and 10 is "Difficult".
    [Collated answers in the results tab.]

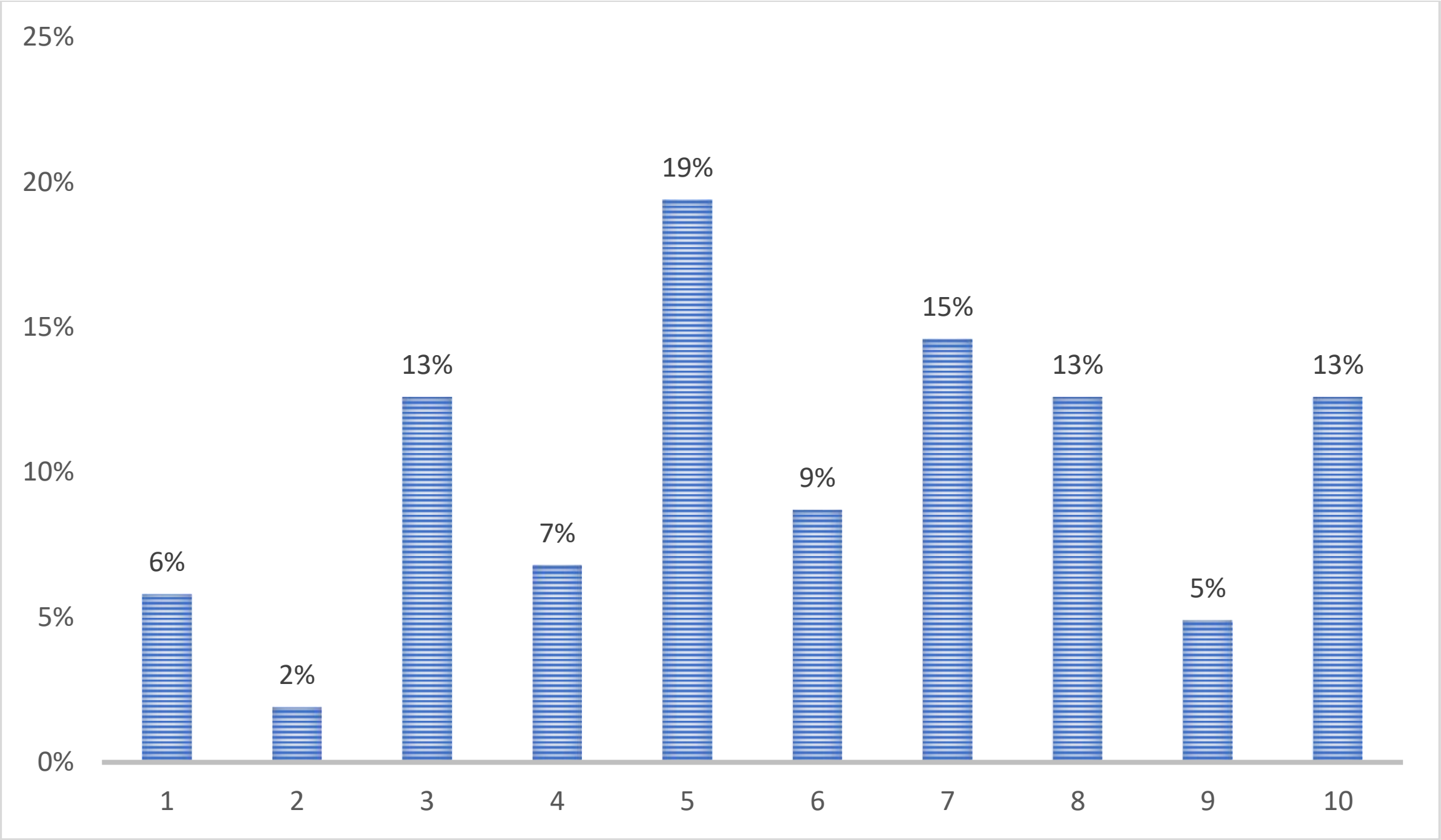

18. How comfortable are you with learning online currently? Please indicate on a scale of 1 to 10 where 1 is "Very Uncomfortable" and 10 is "Very Comfortable".

    [Collated answers in the results tab.]

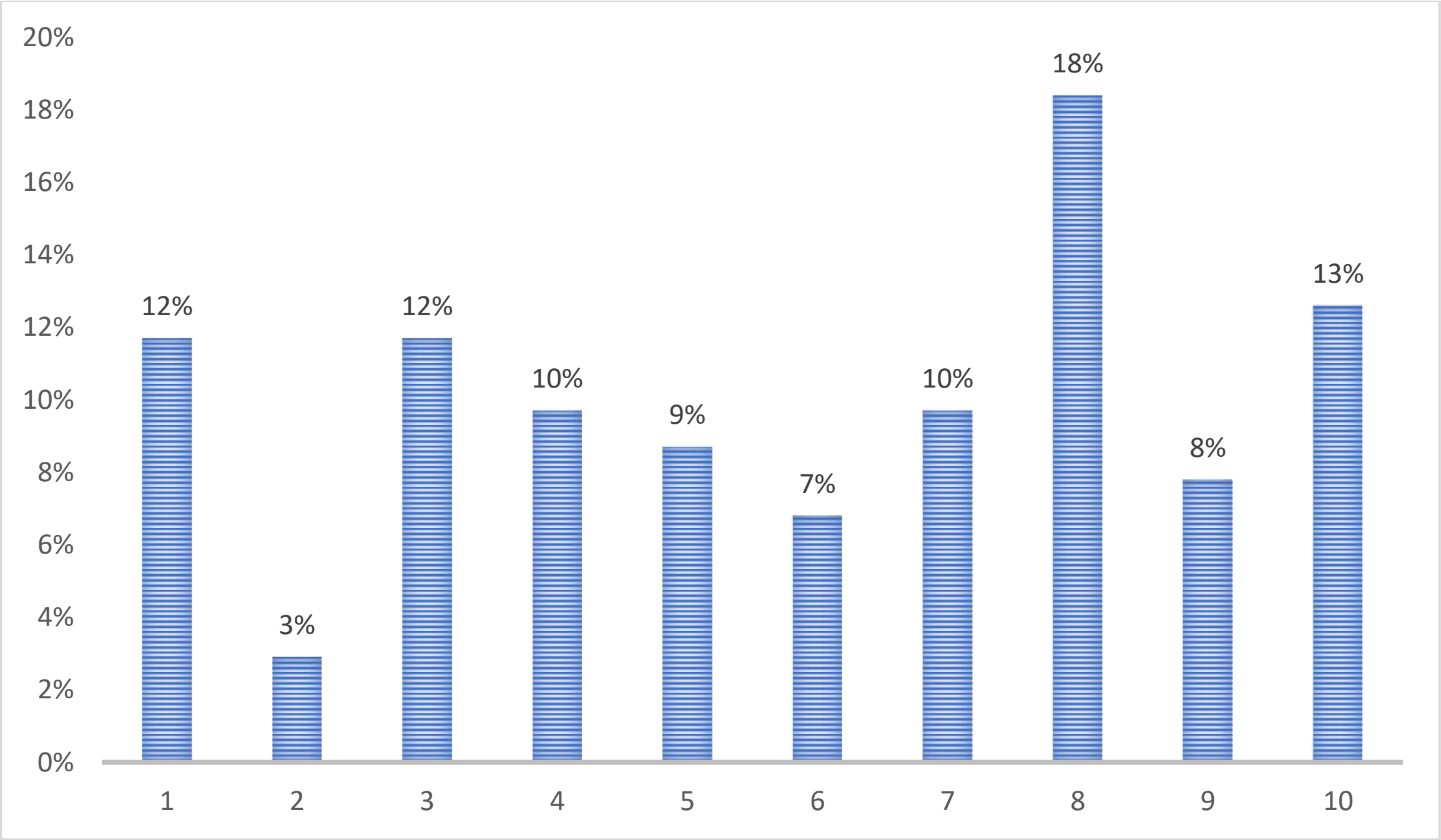

19. How much has your learning style changed, compared to how it used to be pre-COVID-19? Please indicate on a scale of 1 to 10 where 1 is "Not much" and 10 is "A Lot".

[Collated answers in the results tab.]

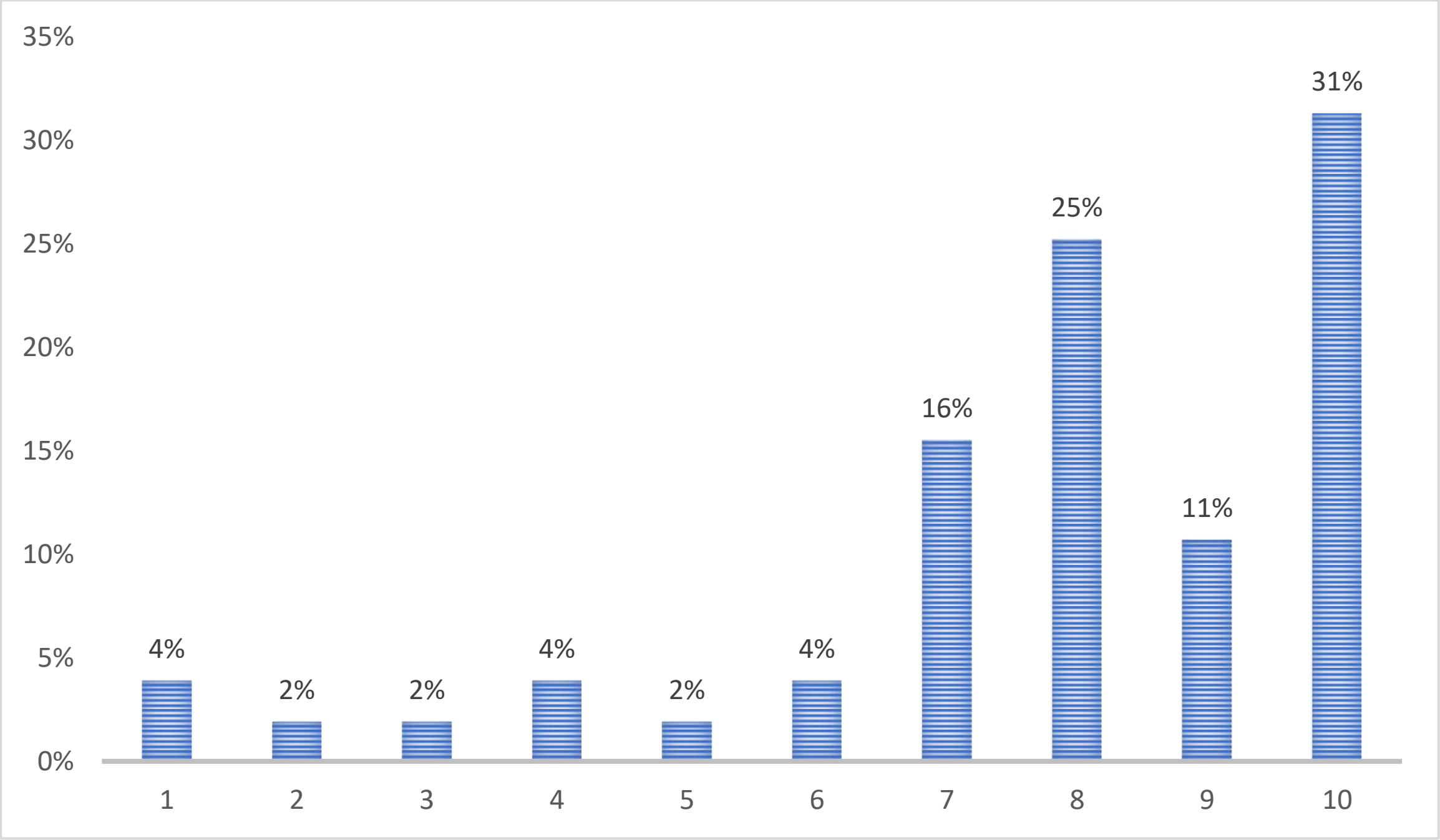

20. Which online educational resources do you use?

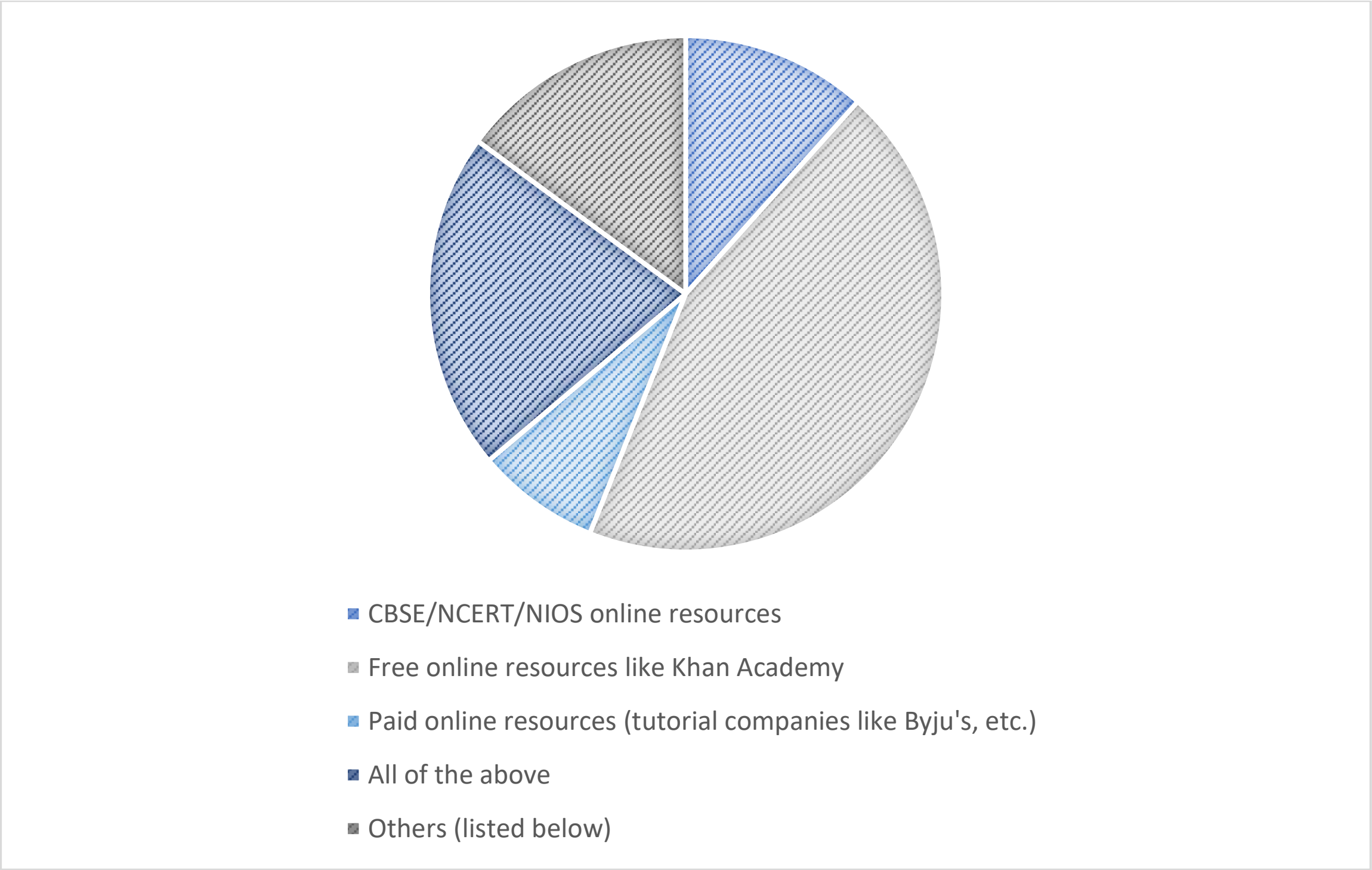

**Others**

**Flashcards**

- Quizlet

**Exam Archives**

- Papa Cambridge
- Save My Exams
- Xtreme Papers

**Videos**

- BBC Bitesize
- YouTube
- SkillsBuild

**Conceptual**

- LitCharts
- Khan Academy

**Miscellaneous**

- Personal tuitions

21. Will you continue to use these resources even after school campuses open?

    [Answers in the results tab.]

22. Do you think you have benefitted from online learning?
    [Answers in the results tab.]

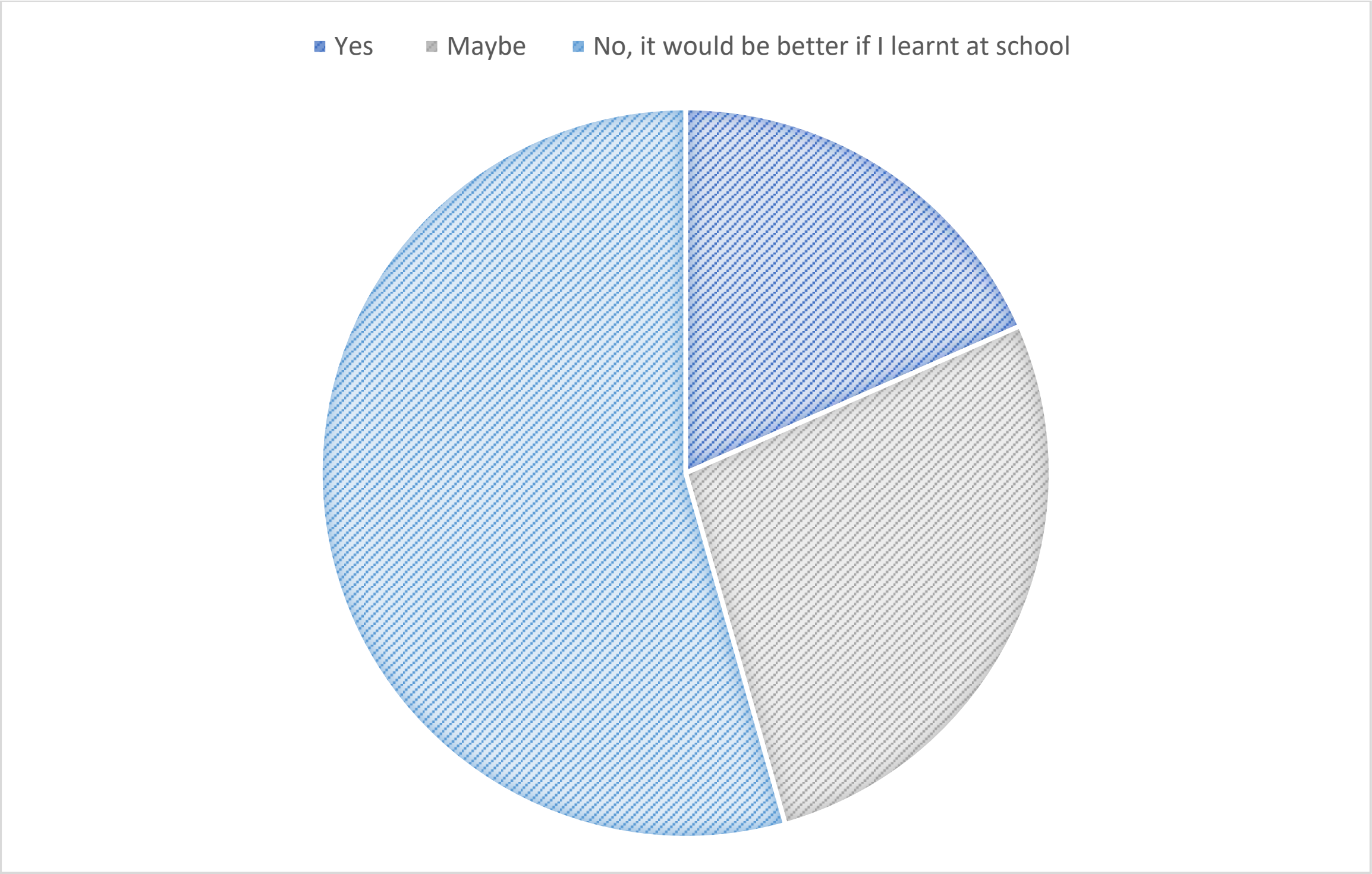

23. If e-learning were not a necessity, would you prefer to learn online, in real life, or a mix of both?

    [Answers in the results tab.]

24. If you chose a mix of both, what ratio of online learning to offline learning would you prefer?
    *Choose the most appropriate answer. If none of the options work for you, please choose 'Other' and write down your preferred percentage.*

    [Answers in the results tab.]

## Section 4: Both Groups

25. How would you rate this survey? Please indicate on a scale of 1 to 10 where 1 is “Very Bad” and 10 is “Excellent”.

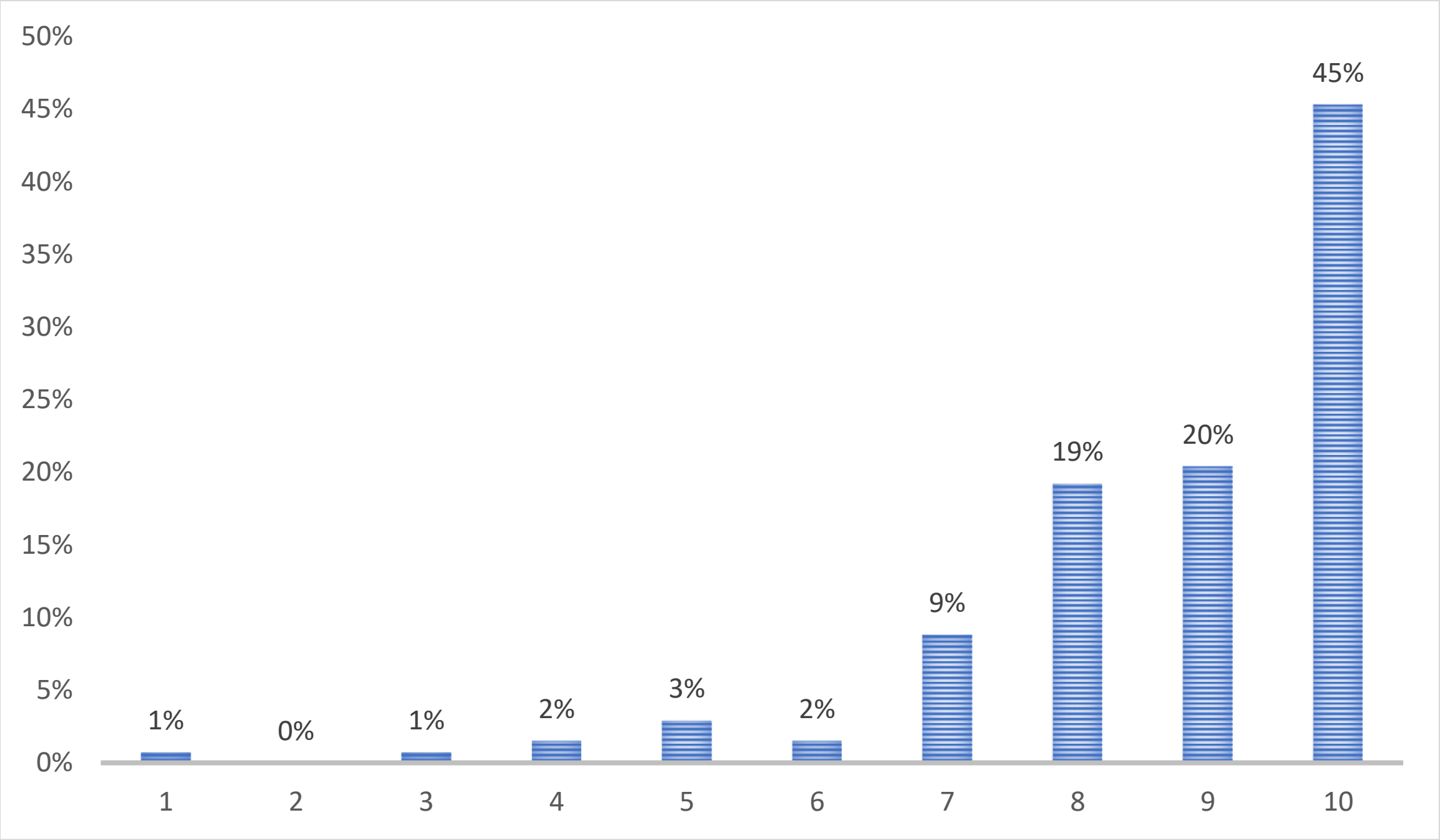

# Annexure-3: About

## Avni Singh

I wrote this report because of my interest in e-learning and how it can impact underserved students.

I am sixteen years old (as of 2021) and am studying in the IB Diploma program at my school in Gurgaon, India. Following the tradition of social work and entrepreneurship in my family, I founded the Society for Inclusive Education, which is a national student-led movement that aims to level the playing field for students of all backgrounds. I also founded Project Neev, which impacted around 50 children at Happy School, Gurgaon.

I am also deeply interested in economics, particularly how it can be used to improve the lives of people around me.

## Professor HS Jamadagni

Professor Jamadagni has been an academic faculty member for over four decades in the Indian Institute of Science in Bangalore, involved in teaching, research, and industry consulting. He was also a member of the Telecom Regulatory Authority of India (2010-13). After superannuation, he continues to be involved in academics.

He guided me in writing this report.